\documentclass[pra,twocolumn,floatfix,showpacs]{revtex4}
\pdfoutput=1
\usepackage{graphicx}
\usepackage{simon}
\usepackage{bbm}

\newcommand{\ot}{\widetilde{\Go}}

\def\clap#1{\hbox to 0pt{\hss#1\hss}}
\def\mathclap{\mathpalette\mathclapinternal}
\def\mathclapinternal#1#2{%
	\clap{$\mathsurround=0pt#1{#2}$}%
}

\begin{document}

\title{Light-induced polarization effects in atoms with partially resolved hyperfine structure
and applications to absorption, fluorescence, and nonlinear
magneto-optical rotation}
\author{M. Auzinsh}
\email{Marcis.Auzins@lu.lv} \affiliation{Department of Physics
and Laser Center, University of Latvia, 19 Rainis boulevard,
Riga LV-1586, Latvia}
\author{D. Budker}
\email{budker@berkeley.edu} \affiliation{Department of Physics,
University of California, Berkeley, CA 94720-7300, USA}
\affiliation{Nuclear Science Division, Lawrence Berkeley
National Laboratory, Berkeley CA 94720, USA}
\author{S. M. Rochester}
\email{simonr@berkeley.edu} \affiliation{Department of Physics,
University of California, Berkeley, CA 94720-7300, USA}

\date{\today }

\begin{abstract}
The creation and detection of atomic polarization is examined
theoretically, through the study of basic optical-pumping
mechanisms and absorption and fluorescence measurements, and
the dependence of these processes on the size of ground- and
excited-state hyperfine splittings is determined. The
consequences of this dependence are studied in more detail for
the case of nonlinear magneto-optical rotation in the Faraday
geometry (an effect requiring the creation and detection of
rank-two polarization in the ground state) with alkali atoms.
Analytic formulas for the optical rotation signal under various
experimental conditions are presented.
\end{abstract}


\maketitle

\section{Introduction}

Since the pioneering work of Alfred Kastler and Jean Brossel in
the 1950s \cite{Kas67}, atomic polarization created by the
interaction of light with atoms have been an exciting topic of
research, providing new methods for laser spectroscopy and
delivering new technologies for practical applications, such as
narrow-band optical filters \cite{Yeh82}.

Atomic polarization, created in a medium by polarized light,
can modify the optical response of the medium, affecting the
light field. For example, the absorption of light of a
particular polarization by atoms in a polarized state can be
reduced (electromagnetically induced transparency \cite{Fle05})
or increased (electromagnetically induced absorption
\cite{Lez99}) compared to that for an unpolarized state.
Coherent population trapping \cite{Nas06} is a closely related
phenomenon, the study of which led to the discovery of an
interesting effect that is also a powerful tool for the
manipulation of atomic states: coherent population transfer
between atomic states, known as STIRAP (stimulated Raman
adiabatic passage) \cite{Ber98}. ``Lasing without inversion''
\cite{Koc88,Scu89} is another related effect.

Additional effects are encountered when atoms interact with
coherent light in the presence of a magnetic field
\cite{Bud2002RMP,Ale2005}. (Reference \cite{Bud04Lambda}
discusses a relationship between these effects and
electromagnetically induced absorption.) These magneto-optical
effects---especially those involving magnetic-field-induced
evolution of long-lived ground-state polarization---can be used
to perform sensitive magnetometery \cite{Bud2007NatPhys}.
(These effects are also often referred to as ``coherence
effects'', although this is something of a misnomer, as in some
cases the effects can be described using a basis in which there
are no ground-state coherences \cite{Kan93}.)

The atomic polarization responsible for specific effects, such
as nonlinear magneto-optical rotation (NMOR), can be described
in terms of the polarization moments (PM) in the multipole
expansion of the density matrix \cite{Omo77,Auz95}. The
lowest-rank multipole moments correspond to population,
described by a rank $\kappa=0$ tensor, orientation, described
by a rank $\kappa=1$ tensor, and alignment, described by a rank
$\kappa=2$ tensor. It is these three lowest-rank multipole
moments that can directly affect light absorption and
laser-induced fluorescence \cite{Dya65,Auz95}, and thus can be
created and detected through single-photon interactions. An
atomic state with total angular momentum $F$ can support
multipole moments with rank up to $\Gk=2F$ \cite{Omo77,Auz95};
multi-photon interactions and multipole transitions higher than
dipole allow the higher-order moments to be created and
detected. Magneto-optical techniques can be used to selectively
address individual high-rank multipoles
\cite{Auz84,Yas2003Select}. Recently, the possibility of using
the $\kappa=4$ hexadecapole moment to improve the
characteristics of atomic magnetometers was studied (see
Ref.~\cite{Aco08} and references therein). Effects due to the
$\kappa=6$ hexacontatetrapole moment have also been observed
\cite{Auz83,Pus06}.

Magneto-optical coherence effects that involve linearly
polarized light generally require the production and detection
of polarization corresponding to atomic alignment. (There are
multi-field, high-light-power effects in which alignment is
converted to orientation, which is then detected
\cite{Auz92,Bud2000AOC,Auz2006}; these effects still depend on
the creation of alignment by the light.) Thus, for ground-state
coherence effects, the ground state in question must have
angular momentum of at least $F=1$ in order to support a
rank-two polarization moment. The alkali atoms K, Rb, and
Cs---commonly used for magneto-optical experiments---each have
ground-state hyperfine sublevels with $F\ge1$. If light is
tuned to a suitable transition between a ground-state and an
excited-state hyperfine sublevel, alignment can be created and
detected in the ground state.

The situation changes, however, if the hyperfine structure is
not resolved. If the hyperfine transitions are completely
unresolved (as was the case in early work that used broadband
light sources such as electrodeless discharge lamps to excite
atoms), then it is the fine-structure transition that is
effectively excited---the $D1$ line ($n^{2}S_{1/2}\rightarrow
n^{2}P_{1/2}$) or the $D2$ line ($n^{2}S_{1/2}\rightarrow
n^{2}P_{3/2}$). In this case, the effects related to the
excitation of a particular hyperfine transition are averaged
out when all transitions are summed over. Thus the effect of
the nuclear spin is removed, and the states have effective
total angular momentum $J=1/2$ for the ground state and $J=1/2$
or $3/2$ for the excited state. In this case the highest rank
multipole moment that can be supported by the ground state is
orientation ($\kappa=2J=1$), and effects depending on atomic
ground-state alignment will not be apparent.

In practical experiments with alkali atoms in vapor cells, even
when narrow-band laser excitation is used, the hyperfine
structure is in general only partially resolved, due to Doppler
broadening. At room temperature, the Doppler widths of the
atomic transitions in K, Rb, and Cs range from 463 MHz for K to
226 MHz for Cs. The ground-state hyperfine splittings, ranging
from 462 MHz for K to 9.192 GHz for Cs, are on the order of or
greater than the Doppler widths, while the excited-state
hyperfine splittings, ranging from 8 MHz to 1.167 GHz, are
generally on the order of or smaller than the Doppler width.
Thus the question arises: how do coherence effects depend on
the ground- and excited-state hyperfine splitting when the
hyperfine structure is neither completely resolved nor
completely unresolved?

In Sec.~\ref{sec:unresolved} we discuss transitions for which
one or the other of the excited- or ground-state hfs is
completely unresolved. We determine which polarization moments
can be created in the ground state via single-photon
interactions, and which moments can be detected through their
influence on light absorption. We find that the two
contributions to the ground-state polarization---absorption and
polarization transfer through spontaneous decay---depend
differently on the ground- and excited-state hyperfine
structure.

In Sec.~\ref{sec:nmoe}, we choose a particular system (the D1
and D2 lines of alkali atoms) and investigate the detailed
dependence of NMOR signals on the excited- and ground-state
hyperfine splitting. We consider three cases: systems in which
the atomic Doppler distribution can be neglected, and systems
in which the Doppler distribution is broad compared to the
natural line width and in which the rate of velocity-changing
collisions is either much slower than or much faster than the
ground-state polarization relaxation rate. Appendix
\ref{app:NMORcalc} contains some general results used in
Sec.~\ref{sec:nmoe} and some more details of the calculation.

Throughout the discussion we use the low-light-intensity
approximation in order to simplify the calculations and obtain
analytic results. It can be shown, using higher-order
perturbation theory and numerical calculations, that the
essential results presented here hold for arbitrary light
intensity, as well. Previous work that discusses the dependence
of optical pumping on whether or not hfs is resolved includes
Refs.~\cite{Hap67b,Mat70,Hap72,Leh67,Gaw2002}.

\section{Totally unresolved ground- or excited-state hyperfine structure}
\label{sec:unresolved}

In this section, we discuss the creation and detection of
atomic polarization in systems for which either the ground- or
excited-state hyperfine structure is unresolved. This section
deals with systems that can be described using the
complete-mixing approximation, i.e., the assumption that atomic
velocities are completely rethermalized in between optical
pumping and probing. This is the case for experiments using
buffer-gas or antirelaxation-coated vapor cells, in which atoms
undergo frequent velocity-changing collisions during the
ground-state polarization lifetime. The consequences of the
complete mixing approximation are similar to those of the
broad-line approximation \cite{Bar61a,Bar61b,Coh62I,Coh62II},
which takes the spectrum of the pump light to be broader than
the Doppler width of the ensemble. In the complete-mixing case,
narrow-band light produces polarization in a single velocity
group in the Doppler distribution, and the polarization is
averaged over all velocity groups through rethermalization,
while in the broad-line case, the entire Doppler distribution
is pumped directly.

\subsection{Depopulation pumping}
\label{sec:depop}

We consider an ensemble of atoms with nuclear spin $I$, a
ground state with electronic angular momentum $J_g$, and an
excited state with angular momentum $J_e$. The various ground-
and excited-state hyperfine levels are labeled by $F_g$ and
$F_e$, respectively. The atoms are subject to weak
monochromatic light with complex polarization vector $\uv{e}$
and frequency $\Go$, near-resonant with the atomic transition
frequency $\Go_{J_gJ_e}$. We assume that the atoms undergo
collisions that mix different components of the Doppler
distribution (or, equivalently, use the broad-line
approximation). We also neglect coherences between different
ground-state or different excited-state hyperfine levels (these
coherences will not develop for low light power as long as the
hyperfine splittings are larger than the natural width of the
excited state). We first consider polarization produced in the
ground state due to atoms absorbing light and being transferred
to the excited state (depopulation pumping). The general form
of the contribution to the ground-state density matrix due to
this effect is given by \cite{Hap72} (see also
Refs.~\cite{Bar61a,Bar61b,Coh62I,Coh62II} for the derivation in
the context of the broad-line approximation)
\begin{equation}\label{eq:depopgeneral}
    \Gr^{(depop)}_{mn}
    \propto\sum_r\uv{e}^\ast\cdot\V{d}_{mr}
        \uv{e}\cdot\V{d}_{rn}G(\Go-\Go_{rn}),
\end{equation}
where $m$ and $n$ are degenerate ground states, $r$ is an
excited state, $\Go_{rn}$ is the transition frequency between
$r$ and $n$, and $G$ is a function describing the spectral line
shape. If the natural width of the excited state is much
smaller than the Doppler width $\GG_D$, $G$ is approximately a
Gaussian of the Doppler width. For the system described above,
this takes the form
\begin{equation}\label{eq:rhoabs}
\begin{split}
    \Gr^{(depop)}_{F_gm,F_gm'}\propto
    &{}\sum_{\mathclap{F_em''}}\bra{F_gm}\uv{e}^\ast\cdot\V{d}\ket{F_em''}\\
    &\qquad
    \times\bra{F_em''}\uv{e}\cdot\V{d}\ket{F_gm'}\,
    G(\Go-\Go_{F_eF_g}).
\end{split}
\end{equation}

\begin{figure}
    \includegraphics{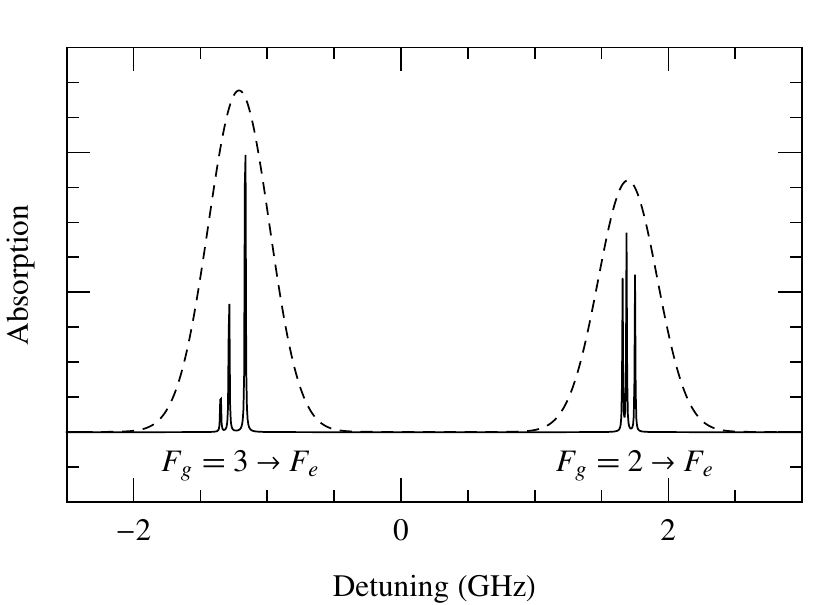}
    \caption{Doppler-free (solid line) and Doppler-broadened
    (dashed line) absorption spectra for the $^{85}$Rb D2 line.
    A Maxwellian velocity distribution at room temperature is
    assumed. If the incident light frequency is tuned near the
    center of the $F_g=2\rightarrow F_e$ transition group, the
    condition discussed in the text is fulfilled. Namely, the
    light detuning from each resonance frequency is either much
    less than or much greater than the Doppler width. The
    condition holds somewhat less rigorously for light tuned to
    the center of the $F_g=3\rightarrow F_e$ transition group.}
    \label{fig:rb87D2spectra}
\end{figure}

Now suppose that the light frequency is tuned so that it is
close, compared with the Doppler width, to an unresolved group
of transition frequencies, and far from every other transition
frequency (Fig.~\ref{fig:rb87D2spectra}).
We employ the simplest approximation that $G(\Go-\Go_{F_eF_g})$
takes the same value for each transition in the unresolved
group, and is zero for all other transitions. With these
approximations, Eq.~\eqref{eq:rhoabs} becomes
\begin{equation}\label{eq:rhoabsapprox}
\begin{split}
    \Gr^{(depop)}_{F_gm,F_gm'}\propto
    \sum_{\mathclap{F_em''}}\bra{F_gm}\uv{e}^\ast\cdot\V{d}\ket{F_em''}
    \bra{F_em''}\uv{e}\cdot\V{d}\ket{F_gm'},
\end{split}
\end{equation}
where the sum now runs over only those excited states $F_e$
that connect via one of the unresolved resonant transitions to
the ground state $F_g$ in question. (This sum also arises in
the broad-line approximation.)

We now investigate which coherences can be created in the
ground state by the light. As we will see, this will determine
which polarization moments can be created. Suppose first that
the excited-state hfs is entirely unresolved. Then the sum over
$\ket{F_em''}\bra{F_em''}$ in Eq.~\eqref{eq:rhoabsapprox} runs
over all excited states, so that it is equivalent to the
identity. We replace this sum with the sum over the eigenstates
in the uncoupled basis
$\sum_{m_I''m_J''}\ket{Im_I''J_em_J''}\bra{Im_I''J_em_J''}$.
Further, we insert additional sums to expand the ground-state
coupled-basis eigenstates in terms of the uncoupled basis. We
also expand $\uv{e}$ and $\V{d}$ in terms of their spherical
components. Equation \eqref{eq:rhoabsapprox} becomes
\begin{widetext}
\begin{equation}\label{eq:coupledunresolved}
\begin{split}
    \Gr^{(depop)}_{F_gm,F_gm'}
    &\propto\sum(-1)^{q'+q''}(e^\ast)_{q'}e_{q''}
    \braket{F_gm}{Im_IJ_gm_J}\bra{Im_IJ_gm_J}d_{-q'}\ket{Im_I''J_em_J''}\\
    &\qquad\qquad\qquad\qquad\qquad\qquad\qquad\qquad\qquad\qquad\qquad
    \times\bra{Im_I''J_em_J''}d_{-q''}\ket{Im_I'J_gm_J'}
    \braket{Im_I'J_gm_J'}{F_gm'}\\
    &=\sum(-1)^{q'+q''}(e^\ast)_{q'}e_{q''}
    \braket{F_gm}{Im_IJ_gm_J}\bra{J_gm_J}d_{-q'}\ket{J_em_J''}
    \bra{J_em_J''}d_{-q''}\ket{J_gm_J'}\braket{Im_IJ_gm_J'}{F_gm'},
\end{split}
\end{equation}
\end{widetext}
where the inner products $\braket{\cdots}{\cdots}$ are given by
the Clebsch-Gordan coefficients, with
$\braket{J_3m_3}{J_1m_1J_2m_2}=\braket{J_1m_1J_2m_2}{J_3m_3}$.
In the second line we have used the fact that the
electric-dipole operator is diagonal in the nuclear-spin
states.

We now use the Clebsch-Gordan condition $m_1+m_2=m_3$, as well
as the related electric-dipole selection rule
\begin{equation}
    \bra{J_1m_1}d_q\ket{J_2m_2}=0\text{ unless }m_1=m_2+q,
\end{equation}
to determine which coherences $\Gr^{(depop)}_{F_gm,F_gm'}$ can
be nonzero in Eq.~\eqref{eq:coupledunresolved}. Traversing the
factors in the last line of Eq.~\eqref{eq:coupledunresolved}
from left to right, we find that a term in the sum is zero
unless
\begin{equation}\label{eq:coupledunresolvedconditions}
\begin{gathered}
    m=m_I+m_J,\ m_J''=m_J+q',\\
    m_J'=m_J''+q'',\ m'=m_I+m_J'.
\end{gathered}
\end{equation}
From this we find that
\begin{equation}
    \abs{m'-m}=\abs{q'+q''}\le2.
\end{equation}
We can translate a limit on $\abs{\GD m}$ directly into a limit
on the rank $\Gk$ of polarization moments that can be created
as follows. The polarization moments are the coefficients of
the expansion of the density matrix into a sum of irreducible
tensor operators (a set of operators with the rotational
symmetries of the spherical harmonics). A polarization moment
of rank $\Gk$ has $2\Gk+1$ components with projections
$q=-\Gk,\ldots,\Gk$, which are related to the Zeeman-basis
density-matrix elements by
\begin{equation}\label{Eq:RhoToKappaQ}
    \Gr^{(\Gk)}_q
    \propto\sum_{m,m'=-F}^F
        \cg{Fm\Gk q}{Fm'}\Gr_{mm'}.
\end{equation}
From Eq.~\eqref{Eq:RhoToKappaQ}, a ground-state PM
$\Gr^{(\Gk)}_q$ with a given value of $\abs{q}$ can exist if
and only if there is a $\abs{\GD m}=\abs{q}$ coherence in the
ground-state density matrix. A limit on $\abs{q}$ is not by
itself a limit on $\Gk$, because any polarization moment with
rank $\Gk\ge\abs{q}$ can have a component with projection $q$.
However, if such a high-rank moment exists, we can always find
a rotated basis such that the component with projection $q$ in
the original basis manifests itself as a component with
projection $\Gk$ in the rotated basis. Because
Eq.~\eqref{eq:coupledunresolved} holds for arbitrary light
polarization, it holds in the rotated basis, so we can conclude
that no polarization moment $\Gr^{(\Gk)}_q$ with rank $\Gk$
greater than the limit on $\abs{\GD m}$ can be created,
regardless of the value of $q$.

For the case under consideration, this analysis reveals that
only polarization moments with $\Gk\le2$ are present. This is a
consequence of the fact that we are considering the
lowest-order contribution to optical pumping (namely, second
order in the incident light field), so that multi-photon
effects are not taken into account. A single photon is a
spin-one particle, so it can support polarization moments up to
$\Gk_\Gg=2$. For a polarization moment of rank $\Gk$ to be
created, the unpolarized (rank 0) density matrix must be
coupled to a rank-$\Gk$ PM by the rank $\Gk_\Gg\le2$ photon.
The triangle condition for tensor products implies that
$\Gk\le\Gk_\Gg+0\le2$.

An additional condition on $\abs{\GD m}$ can be found from Eq.\
\eqref{eq:coupledunresolvedconditions}, using the fact that
$m_J$ and $m_J'$ are projections of the ground-state electronic
angular momentum, so that their absolute values are less than
or equal to $J_g$. From the first and last conditions of Eq.\
\eqref{eq:coupledunresolvedconditions} we find
\begin{equation}
    \abs{m'-m}=\abs{m_J'-m_J}\le2J_g.
\end{equation}
Thus the coherences that can be created within a ground-state
hyperfine level $F_g$ are limited to twice the ground-state
electronic angular momentum $J_g$, even if $F_g>J_g$. As a
consequence, polarization moments in the ground state are
limited to rank $\Gk\le2J_g$. We can understand this
restriction by examining Eq.~\eqref{eq:coupledunresolved}.
Because the excited-state hyperfine shifts have been eliminated
from the expression and the electric dipole operator does not
act on the nuclear spin space, all traces of the hyperfine
interaction in the excited state have been removed from
Eq.~\eqref{eq:coupledunresolved}. This is indicated by the fact
that, in the last line of the equation, the nuclear spin does
not appear in the state vectors describing excited states. Thus
the excited state only couples to the electronic spin of the
ground state, so that there is no mechanism for coupling two
ground-state nuclear spin states. This means that any
polarization moment present in the ground state must be
supported by the electronic spin only.


Considering now the case in which the excited-state hfs is
resolved and the ground-state hfs is unresolved, opposite to
the case considered so far, we find no similar restriction. It
is clear from Eq.~\eqref{eq:rhoabsapprox} that the polarization
produced in a ground-state hyperfine level is independent of
all of the other ground-state levels---only one ground-state
level $F_g$ appears in the equation. If the excited-state hfs
is resolved, then likewise only one excited-state level $F_e$
appears. Thus pumping on a transition $F_g\rightarrow F_e$
produces the same polarization in the level $F_g$ as pumping on
a completely isolated $F_g\rightarrow F_e$ transition,
regardless of any nearby (unresolved) ground-state hyperfine
levels. Any polarization moment up to rank $\Gk=2F_g$ can be
produced, subject to the restriction $\Gk\le2$ in the
lowest-order approximation.

In fact, these results can be obtained without the need for any
calculations. It is clear that if all the hyperfine splittings
are set to zero, the nuclear spin is effectively
noninteracting, and can be ignored. In this case, only
polarization moments that can be supported by the electronic
spin $J_g$ can be produced in the ground state. In particular,
if we consider polarization of a given (degenerate)
ground-state hyperfine level, we must have $\Gk\le2J_g$. If the
ground-state hyperfine splitting is increased, this conclusion
must remain unchanged, because the light only couples the
ground states to the excited states; to lowest order it does
not make any difference what is going on in the other
ground-state hyperfine levels. If the excited-state hyperfine
splitting is then increased, the various $F_g\rightarrow F_e$
hyperfine transitions become isolated; for an isolated
transition the limit on the ground-state polarization moments
is $\Gk\le2F_g$. Thus we see that the limit $\Gk\le2J_g$ on the
ground-state polarization moments occurs when the excited state
hfs is unresolved, and this limit does not depend on whether or
not the ground-state hfs is resolved.

The total angular momentum $F_g$ can be significantly larger
than $J_g$. For example, Cs has $I=7/2$ and $J_g=1/2$, so that
the maximum value of $F_g$ is 4. Thus polarization moments up
to rank eight can be produced in the ground state by
depopulation pumping if the excited-state hfs is resolved, but
only up to rank one if it is unresolved \cite{Mat70}. To second
order in the light field the ground-state polarization that can
be created is limited to at most rank two in any case. However,
the question of whether rank-two polarization can be created is
an important one: ground-state alignment is crucial for
nonlinear magneto-optical effects with linearly polarized
light, as we discuss in Sec.~\ref{sec:nmoe}.

\begin{figure}
    \includegraphics{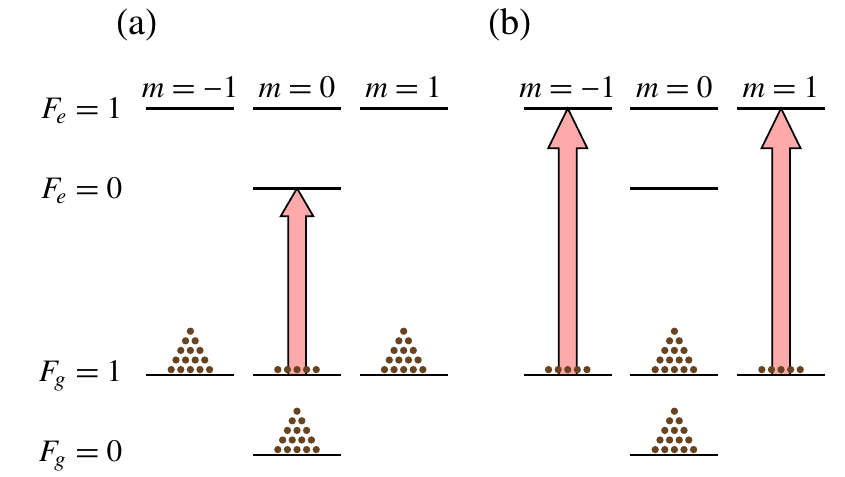}
    \caption{(Color online) Excitation with $z$-polarized light on the (a)
    $F_g=1\rightarrow F_e=0$ and (b) $F_g=1\rightarrow F_e=1$
    transitions of a totally resolved $J_g=1/2\rightarrow
    J_e=1/2$ transition with $I=1/2$. Alignment is produced in
    the $F_g=1$ hyperfine level in both cases. Relative atomic
    populations are indicated by the number of dots displayed
    above each ground-state level. Relative transition
    strengths are indicated by the widths of the arrows---here
    the transition strengths are all the same.}
    \label{leveldiagD1I12_1to0_1to1}
\end{figure}

\begin{figure}
    \includegraphics{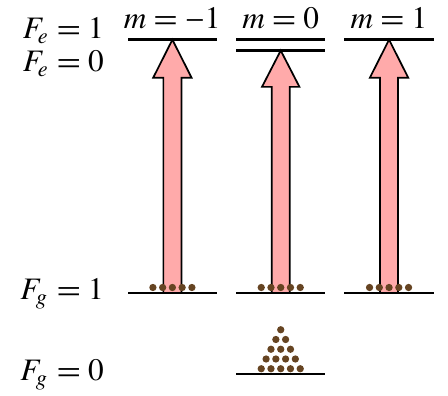}
    \caption{(Color online) As Fig.~\ref{leveldiagD1I12_1to0_1to1} but with excited-state
    hfs unresolved; light is resonant with the $F_g=1\rightarrow F_e$
    transition group. No alignment is produced in the $F_g=1$ ground
    state.} \label{leveldiagD1I12_1to01}
\end{figure}

This situation is illustrated for linearly polarized light
resonant with an alkali D1 line ($J_g=J_e=1/2$) in
Figs.~\ref{leveldiagD1I12_1to0_1to1} and
\ref{leveldiagD1I12_1to01}.
We choose $I=1/2$ for simplicity, and the quantization axis is
taken along the direction of the light polarization. In
Fig.~\ref{leveldiagD1I12_1to0_1to1} the hfs is completely
resolved. Part (a) of the figure shows light resonant with the
$F_g=1\rightarrow F_e=0$ transition. Atoms are pumped out of
the $\ket{F_g=1,m=0}$ sublevel, producing alignment in the
$F_g=1$ state. (Linearly polarized light in the absence of
other fields can only produce even-rank moments, and an $F=1$
state can only support polarization moments up to rank two;
therefore, the anisotropy shown in
Fig.~\ref{leveldiagD1I12_1to0_1to1} must correspond to
alignment.) If light is resonant with the $F_g=1\rightarrow
F_e=1$ transition, as in part (b), the $\ket{F_g=1,m=\pm1}$
sublevels are depleted, producing alignment with sign opposite
to that in Fig.~\ref{leveldiagD1I12_1to0_1to1}a. This can be
contrasted with the case in which the excited-state hyperfine
structure is completely unresolved, shown in
Fig.~\ref{leveldiagD1I12_1to01}. Here, all the Zeeman sublevels
of the $F_g=1$ state are pumped out equally---the
$\ket{F_g=1,m=0}$ sublevel on the $F_g=1\rightarrow F_e=0$
transition, and the $\ket{F_g=1,m=\pm1}$ sublevels on the
$F_g=1\rightarrow F_e=1$ transition. (The relative pumping
rates, which can be found from terms of the sum in
Eq.~\eqref{eq:rhoabsapprox}, are all the same.) Thus no
imbalance is created in the $F_g=1$ sublevel populations, and
no polarization is created in this state.

\begin{figure}
    \includegraphics{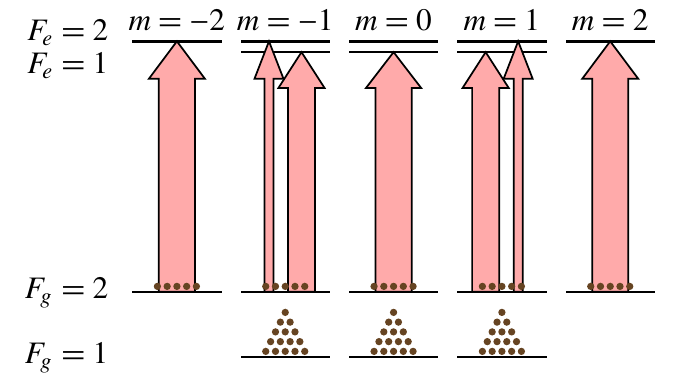}
    \caption{(Color online) Excitation with $z$-polarized light
    on the $F_g=2\rightarrow F_e$ transition group of a $D1$
    transition with unresolved excited-state hfs. The nuclear
    spin is $I=3/2$. No alignment is produced in the $F_g=2$
    hyperfine level. The width of each arrow represents the
    relative transition strength, which can be obtained from
    terms of the sum in Eq.~\eqref{eq:rhoabsapprox}.}
    \label{leveldiagD1I32_2to12}
\end{figure}

The same principle is illustrated for nuclear spin $I=3/2$ in
Fig.~\ref{leveldiagD1I32_2to12}.
The excited-state hfs is unresolved, and light is resonant with
the $F_g=2\rightarrow F_e$ transitions. In this case, the
$m=\pm1$ ground-state sublevels are pumped on two different
transitions. The total transition strength connecting each
$F_g=2$ sublevel to the excited state is the same, and so no
polarization is produced in the $F_g=2$ state.

The conclusions of this section must be modified when
polarization produced in the ground state by spontaneous
emission from the excited state is taken into account. We now
consider the effect of this mechanism on the ground-state
polarization (Sec.~\ref{sec:sepol}).

\subsection{Excited state and repopulation pumping}
\label{sec:sepol}

Through second order in the incident light field (first order
in light intensity), there is one additional contribution to
the ground-state polarization besides the one considered in
Sec.~\ref{sec:depop}: that due to atoms being pumped to the
excited state and then returning to the ground state via
spontaneous emission (repopulation pumping). We first consider
polarization produced in the excited state. The general form of
the excited-state density matrix is \cite{Hap72}
\begin{equation}\label{eq:excitedgeneral}
    \Gr_{rs}
    \propto\sum_k\uv{e}\cdot\V{d}_{rk}
        \uv{e}^\ast\cdot\V{d}_{ks}G(\Go-\Go_{rk}),
\end{equation}
where $r$ and $s$ are excited states and $k$ is a ground state.
Comparing this expression to the formula for ground-state
depopulation pumping [Eq.~\eqref{eq:depopgeneral}], we find
that, as one would expect, the roles of the ground-state and
excited state have been reversed. This means that the results
of Sec.~\ref{sec:depop}, with $F_g$ and $F_e$ interchanged, can
be applied to the excited-state polarization. In this case,
there is a limit $\Gk\le2J_e$ on the polarization moments that
can be produced in the excited state, that occurs only when the
\emph{ground state} hfs is unresolved. The restriction does not
depend on whether or not the excited-state hfs is resolved.
There is the additional limit $\Gk\le2$ for low light power.

When the polarized atoms in the exited state decay due to
spontaneous emission, the polarization can be transferred to
the ground state. This contribution to the ground-state density
matrix is given by \cite{Hap72}
\begin{equation}
    \Gr^{(repop)}_{mn}
    \propto\sum_{sr}\V{d}_{mr}\cdot\V{d}_{sn}\Gr_{rs},
\end{equation}
with $\Gr_{rs}$ as given above. The fact that this formula has
no reference to individual transition frequencies leads us to
expect that the polarization transfer should be independent of
the hyperfine splittings. Indeed, writing this expression out
for the case under consideration gives
\begin{equation}
\begin{split}
    \Gr^{(repop)}_{F_gm,F_gm'}\propto
    &\sum(-1)^{p}
    \bra{F_gm}d_p\ket{F_em''}
    \bra{F_em''}\Gr\ket{F_em'''}\\
    &\qquad\times\bra{F_em'''}d_{-p}\ket{F_gm'},
\end{split}
\end{equation}
and the only restriction to be obtained is $m'-m=m'''-m''$
(excited-state $\GD m$ equals ground-state $\GD m$)
\cite{Coh62I,Coh62II}. (Transforming to the uncoupled basis
does not result in any additional limits.) In other words, if
the polarization moment can be supported in the ground state,
it can be transferred from the excited state via spontaneous
emission.

Combining these results, we see that there is a similar
restriction on polarization created in the ground state by
repopulation pumping as the one on polarization created by
depopulation pumping. However, the restriction occurs in the
opposite case. When the ground state is unresolved the
polarization produced by repopulation pumping must have
$\Gk\le2J_e$. This limit does not depend on whether the
excited-state hfs is resolved.

\begin{figure}
    \includegraphics{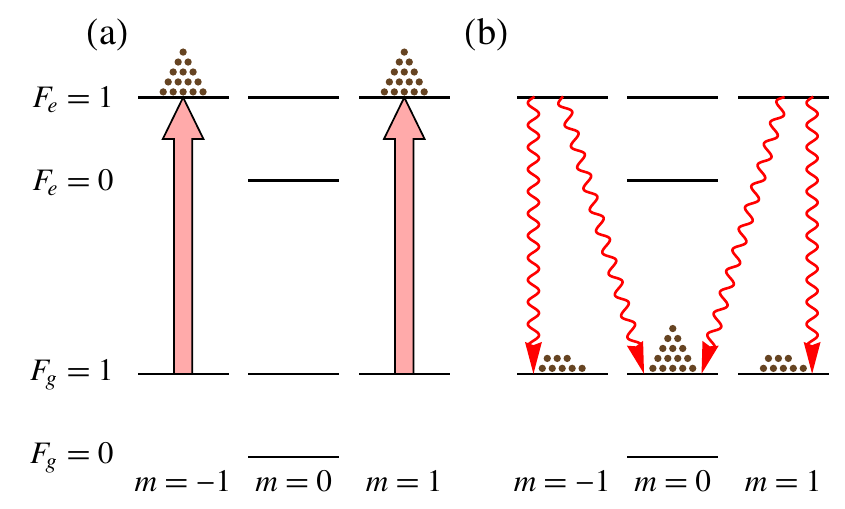}
    \caption{(Color online) Level diagram for an $D1$
    transition with resolved hfs for an atom with $I=1/2$
    showing (a) optical excitation and (b) spontaneous decay
    with linearly polarized light resonant with the
    $F_g=1\rightarrow F_e=1$ transition. The branching ratio
    for each allowed decay is the same, leading to an excess of
    atoms in the $\ket{F_g=1,m=0}$ sublevels over the populations of
    the $\ket{F_g=1,m=\pm1}$ sublevels by a ratio of 2:1.}
    \label{leveldiagD1I12_1to1_SE}
\end{figure}

\begin{figure}
    \includegraphics{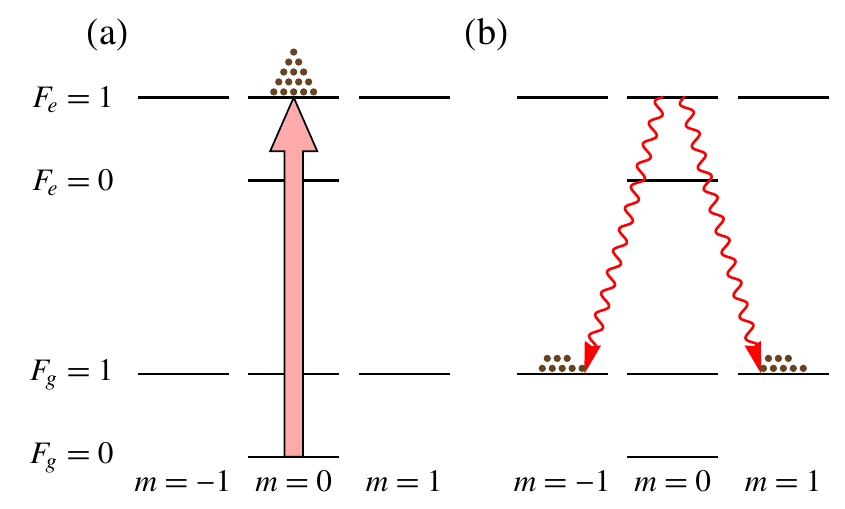}
    \caption{(Color online) As
    Fig.~\ref{leveldiagD1I12_1to1_SE}, but with light tuned to
    the $F_g=0\rightarrow F_e=1$ transition. In this case an
    excess of atoms results in the $m=\pm1$ states, so that the
    polarization has the opposite sign as that in
    Fig.~\ref{leveldiagD1I12_1to1_SE}.}
    \label{leveldiagD1I12_0to1_SE}
\end{figure}

We now illustrate the foregoing for a system with
$J_g=J_e=I=1/2$ pumped with linearly polarized light. In
Fig.~\ref{leveldiagD1I12_1to1_SE} both the ground- and
excited-state hfs is resolved, and light is tuned to the
$F_g=1\rightarrow F_e=1$ transition.
In part (a) of the figure, the pump light produces polarization
in the $F_e=1$ excited state. In part (b) the excited atoms
spontaneously decay. This creates polarization in the $F_g=1$
ground state, because more atoms are transferred to the
$\ket{F_g=1,m=0}$ sublevel than to the $\ket{F_g=1,m=\pm1}$
sublevels. (In this and the following two figures, we do not
show the atoms that decay to the $F_g=0$ state.)
Figure~\ref{leveldiagD1I12_0to1_SE} is the same but with light
tuned to the $F_g=0\rightarrow F_e=1$ transition; polarization
is also created in the $F_g=1$ ground state in this case.

\begin{figure}
    \includegraphics{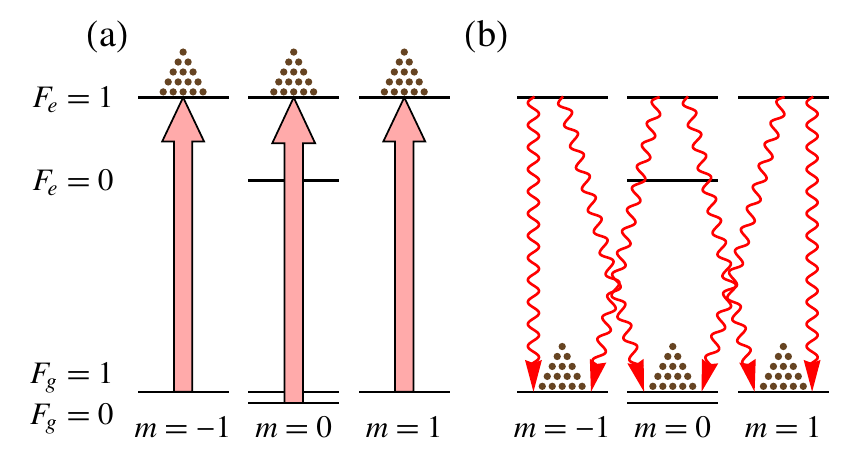}
    \caption{(Color online) As
    Fig.~\ref{leveldiagD1I12_1to1_SE} but with unresolved
    ground-state hfs; light is tuned to the $F_g\rightarrow
    F_e=1$ transition group. The contributions to the
    ground-state polarization illustrated in
    Figs.~\ref{leveldiagD1I12_1to1_SE} and
    \ref{leveldiagD1I12_0to1_SE} cancel, so that no
    ground-state polarization is produced. }
    \label{leveldiagD1I12_01to1_SE}
\end{figure}

In Fig.~\ref{leveldiagD1I12_01to1_SE} the ground-state hfs is
now unresolved, while the excited-state hfs remains resolved.
In this case, both ground-state hyperfine levels are pumped by
the light, and equal populations are produced in the sublevels
of the $F_e=1$ state, as shown in part (a) of the figure. As
seen in part (b), the excited-state atoms spontaneously decay
in equal numbers to the $F_g=1$ sublevels, so that no
polarization is produced in the $F_g=1$ state.

Note that in the opposite case, with unresolved excited-state
hfs and resolved ground-state hfs, spontaneous decay is not
prevented from producing polarization in the $F_g=1$ ground
state.
\begin{figure}
    \includegraphics{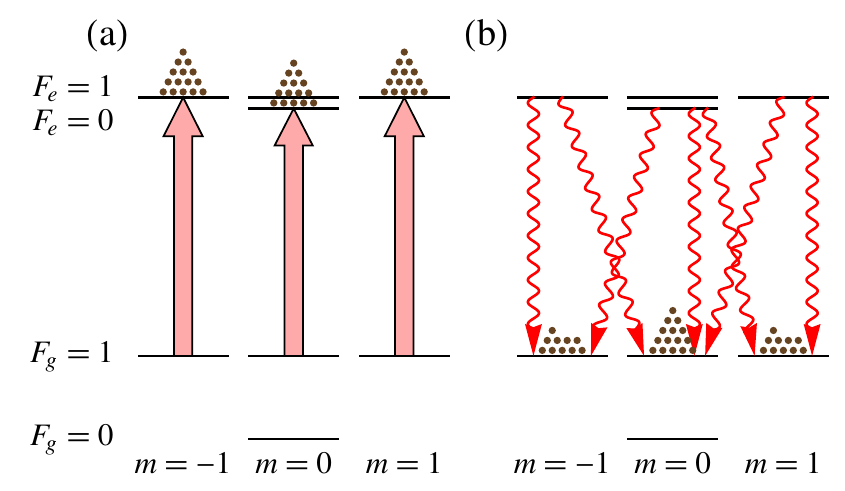}
    \caption{As Fig.~\ref{leveldiagD1I12_1to1_SE} but with
    unresolved excited-state hfs; light is tuned to the
    $F_g=1\rightarrow F_e$ transition group. Three decay
    channels transfer atoms to the $\ket{F_g=1,m=0}$ sublevels, while
    the $\ket{F_g=1,m=\pm1}$ sublevels are each fed by two decay
    channels. Since all the branching ratios are the same, the
    resulting population imbalance is 3:2.}
    \label{leveldiagD1I12_1to01_SE}
\end{figure}
In this case, atoms are pumped into the $F_e=0$ state, as well
as the $F_e=1$ state, as shown in
Fig.~\ref{leveldiagD1I12_1to01_SE}a. Since the
$\ket{F_e=0,m=0}$ state decays isotropically, the decay from
this state does not cancel out the polarization created by
decay from the $F_g=1$ state
(Fig.~\ref{leveldiagD1I12_1to01_SE}b). Thus we see that it is
the ground-state hfs, and not the excited-state hfs, that needs
to be resolved in order for polarization to be produced in the
ground state due to spontaneous decay.

To summarize the results obtained so far, to lowest order in
the excitation light, polarization can be either produced in
the ground state directly through absorption, or transferred to
the ground state by spontaneous emission. To this order,
polarization moments due to both of these mechanisms must have
rank $\Gk\le2$. In addition, if the excited-state hfs is
unresolved, there is a limit $\Gk\le2J_g$ on the ground-state
polarization due to depopulation, but no additional limit on
the polarization due to repopulation. On the other hand, if the
ground-state hfs is unresolved, there is a limit $\Gk\le2J_e$
on the ground-state polarization due to repopulation, but no
additional limit on polarization due to depopulation. Thus,
unless both the excited-state and ground-state hyperfine
structure is unresolved, one or the other of the mechanisms is
capable of producing polarization of all ranks $\Gk\le2$.

\subsection{Light absorption}
\label{sec:abs}

The absorption $\mathcal{A}$ of a weak probe light beam is
given in terms of the ground-state density matrix by
\cite{Hap72}
\begin{equation}
    \mc{A}
    \propto\sum_{mnr}\uv{e}\cdot\V{d}_{rm}\Gr_{mn}
        \uv{e}^\ast\cdot\V{d}_{nr}G(\Go-\Go_{rm}),
\end{equation}
or
\begin{widetext}
\begin{equation}
\begin{split}
    \mathcal{A}\propto
    \sum
    \bra{F_em}\uv{e}\cdot\V{d}\ket{F_gm'}
    \bra{F_gm'}\Gr\ket{F_gm''}
    \bra{F_gm''}\V{d}\cdot\uv{e}^\ast\ket{F_em}
    G(\Go-\Go_{F_eF_g}),
\end{split}
\end{equation}
where all quantities are as defined above. Using the
approximation, as in Secs.\ \ref{sec:depop} and
\ref{sec:sepol}, that the light is resonant with an unresolved
transition group and far detuned from all other transitions,
this formula reduces to
\begin{equation}\label{eq:absformula}
\begin{split}
    \mathcal{A}\propto
    \sum
    \bra{F_em}\uv{e}\cdot\V{d}\ket{F_gm'}
    \bra{F_gm'}\Gr\ket{F_gm''}
    \bra{F_gm''}\V{d}\cdot\uv{e}^\ast\ket{F_em},
\end{split}
\end{equation}
\end{widetext}
where the sum over $F_g$ and $F_e$ includes only those
combinations that are in the unresolved resonant transition
group.

We now investigate the dependence of the absorption on the
ground-state polarization in various cases. Consider the case
in which the ground-state hfs is completely resolved, and the
excited-state structure is unresolved. The light is tuned to a
unresolved transition group consisting of transitions between
one ground-state hyperfine level $F_g$ and all of the
excited-state levels. The sum in Eq.~\eqref{eq:absformula} over
the excited states is then a closure relation, and can be
replaced with a sum over any complete basis for the excited
state, in particular, the uncoupled basis. We also insert
closure relations to expand the ground states $\bra{F_gm''}$
and $\ket{F_gm'}$ in the uncoupled basis. We obtain
\begin{widetext}
\begin{equation}
\begin{split}
    \mathcal{A}
    &\propto
    \sum(-1)^{q'+q''}e_{q'}(e^\ast)_{q''}
    \bra{Im_IJ_em_J}d_{-q'}\ket{Im_I'J_gm_J'}
    \braket{Im_I'J_gm_J'}{F_gm'}\bra{F_gm'}\Gr\ket{F_gm''}
    \braket{F_gm''}{Im_I''J_gm_J''}\\
    &\qquad\qquad\qquad\qquad\qquad
    \times\bra{Im_I''J_gm_J''}d_{-q''}\ket{Im_IJ_em_J}
    \\
    &=
    \sum(-1)^{q'+q''}e_{q'}(e^\ast)_{q''}
    \bra{J_em_J}d_{-q'}\ket{J_gm_J'}\braket{Im_IJ_gm_J'}{F_gm'}
    \bra{F_gm'}\Gr\ket{F_gm''}
    \braket{F_gm''}{Im_IJ_gm_J''}\bra{J_gm_J''}d_{-q''}\ket{J_em_J}
    ,
\end{split}
\end{equation}
\end{widetext}
where only one nuclear-spin summation variable remains in the
last line. The dipole matrix element selection rules and
Clebsch-Gordan conditions require that
\begin{equation}\label{eq:absconditions}
\begin{gathered}
    m'=m_I+m'_J,\
    m'_J=m_J+q',\\
    m''=m_I+m''_J,\
    m_J=m''_J+q'',
\end{gathered}
\end{equation}
must be satisfied in order for a term in the sum to contribute
to the absorption. These conditions can be combined to yield
$\abs{m'-m''}=\abs{q'+q''}\le2$. Thus only coherences with
$\abs{\GD m}\le2$ (and polarization moments with $\Gk\le2$) can
affect the lowest-order absorption signal. The reason for this
is analogous to the reason that polarization moments of maximum
rank two can be created with a lowest-order interaction with
the light. Absorption occurs when an atom is transferred to the
excited state, i.e., when population (rank zero polarization)
is created in the excited state. Thus, to be observed in the
signal, a ground-state atomic PM must be coupled to a $\Gk=0$
excited-state PM by a spin-one photon, which can support
polarization moments up to rank two. The triangle condition for
tensor products then implies that the rank of the atomic
polarization moment must be no greater than two.

Another restriction on the coherences that can affect
absorption can be found from Eq.~\eqref{eq:absconditions} by
using the fact that $|m_J'|\le J_g$ and $|m_J''|\le J_g$. We
find
\begin{equation}
    \abs{m'-m''}=\abs{m'_J-m''_j}\le2J_g.
\end{equation}
In other words, only polarization moments with $\Gk\le2J_g$ can
affect the absorption signal, regardless of the value of $F_g$.
Evidently, it is the excited-state hfs that determines which
ground-state polarization moments can be detected in
absorption, whether or not the ground-state hfs is resolved.

Considering the case in which both the excited- and
ground-state hfs is entirely unresolved can lend some insight
into this result. In this case, every combination of $F_g$ and
$F_e$ enters in the sum in Eq.~\eqref{eq:absformula}. If the
ground-state hyperfine splitting is sent to zero, the sum must
be extended to include matrix elements of $\Gr$ between
different hyperfine levels. This means that all of the sums in
Eq.~\eqref{eq:absformula} can be replaced with sums over
uncoupled basis states, giving
\begin{widetext}
\begin{equation}
\begin{split}
    \mathcal{A}
    &\propto
    \sum(-1)^{q'+q''}e_{q'}(e^\ast)_{q''}
    \bra{Im_IJ_em_J}d_{-q'}\ket{Im_I'J_gm_J'}
    \bra{Im_I'J_gm_J'}\Gr\ket{Im_I''J_gm_J''}
    \bra{Im_I''J_gm_J''}d_{-q''}\ket{Im_IJ_em_J}\\
    &=
    \sum(-1)^{q'+q''}e_{q'}(e^\ast)_{q''}
    \bra{J_em_J}d_{-q'}\ket{J_gm_J'}
    \bra{Im_IJ_gm_J'}\Gr\ket{Im_IJ_gm_J''}
    \bra{J_gm_J''}d_{-q''}\ket{J_em_J}.
\end{split}
\end{equation}
\end{widetext}
Since the hyperfine interaction has been effectively
eliminated, the absorption no longer depends on the nuclear
spin: the complete density matrix does not enter, but rather
the reduced density matrix
\begin{equation}
    \Gr^{(J)}_{m_J'm_J''}=\sum_{m_I}\Gr_{m_Im_J',m_Im_J''}
\end{equation}
that is averaged over the nuclear spin $m_I$. The reduced
density matrix can only support polarization moments up to rank
$\Gk=2J_g$, so any PM in $\Gr$ with higher rank cannot affect
the absorption. Considering a density matrix that is nonzero
only within one ground-state hyperfine level $F_g$, we see that
polarization moments with rank greater than two will not
contribute to the signal. Since the other ground-state
hyperfine levels are unoccupied, it makes no difference what
the ground-state hyperfine splitting is, so we regain the
result that, even if the ground-state hfs is resolved, only
polarization moments with $\Gk\le2J_g$ can affect the
absorption of light if the excited-state hfs is unresolved.

There is no corresponding restriction on the polarization
moments that can affect absorption when the ground-state hfs is
unresolved and the excited-state hfs is resolved. Indeed, we
can consider the case in which only one ground-state hyperfine
level $F_g$ is populated: the absorption is then exactly as if
the transition $F_g\rightarrow F_e$ were completely isolated.
For such an isolated transition, the only limit on detectable
polarization moments is $\Gk\le2$ for the low-power case.

\begin{figure}
    \includegraphics{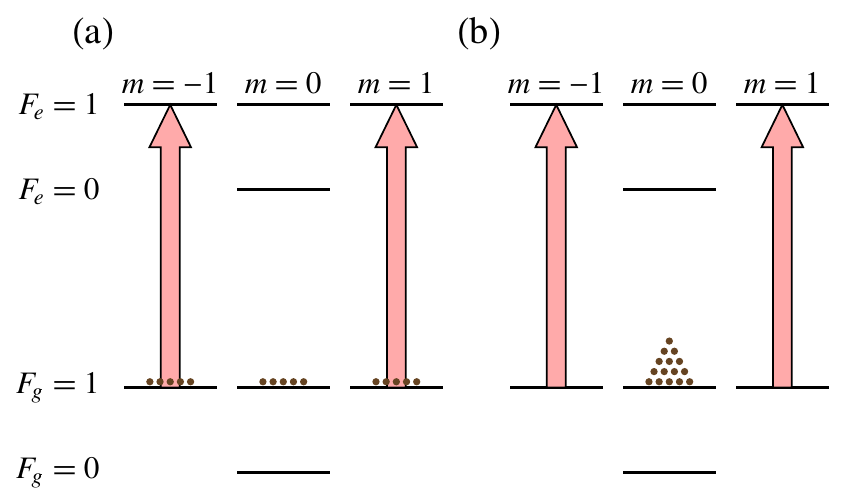}
    \caption{(Color online) $D1$ transition for an atom with $I=1/2$
    subject to linearly polarized light resonant with the
    $F_g=1\rightarrow F_e=1$ transition. In part (a) the
    $F_g=1$ ground state is unpolarized and there is light
    absorption. In part (b) the $F_g=1$ state has the same
    total population, but is aligned, and there is no
    absorption.} \label{leveldiagD1I12_1to1_abs}
\end{figure}

As in the previous subsections, we illustrate this result for a
$D1$ transition for an atom with $I=1/2$ subject to linearly
polarized light. In Fig.~\ref{leveldiagD1I12_1to1_abs} both the
ground- and excited-state hfs is resolved, and the light is
resonant with the $F_g=1\rightarrow F_e=1$ transition.
In part (a) there is no polarization in the $F_g=1$ ground
state: atoms are equally distributed among the Zeeman
sublevels. Light is absorbed by atoms in the
$\ket{F_g=1,m=\pm1}$ sublevels. In part (b) there are the same
total number of atoms in the $F_g=1$ state, but they are
collected in the $m=0$ sublevel. The population is the same,
but the $F_g=1$ state now also has alignment. In this
particular case there is no absorption, because the atoms are
all in the $m=0$ dark state. Thus, in this situation, the
rank-two polarization moment has a strong effect on the
absorption signal.

\begin{figure}
    \includegraphics{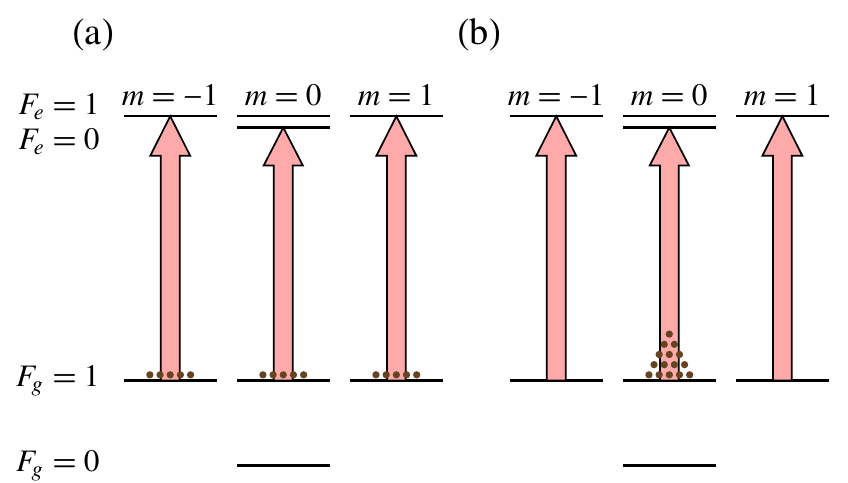}
    \caption{(Color online) As Fig.~\ref{leveldiagD1I12_1to1_abs}, but with
    unresolved excited-state hfs. In this case there is no difference in
    the absorption seen for an (a) unpolarized and (b) aligned $F_g=1$
    ground state.} \label{leveldiagD1I12_1to01_abs}
\end{figure}

Figure \ref{leveldiagD1I12_1to01_abs} shows the same system,
but with unresolved excited-state hfs.
In this case there is no dark state; all of the atoms interact
with the light. The distribution of the atoms among the Zeeman
sublevels does not affect the light absorption, and so the rank
two polarization moment is not detectable in the absorption
signal.

\subsection{Fluorescence}

Finally, we consider which excited-state polarization moments
can be observed in fluorescence. Assuming broadband detection,
the intensity of fluorescence into a particular polarization
$\uv{e}$ is given in terms of the excited-state density matrix
by
\begin{equation}\label{eq:fluorescencegeneral}
    \mc{I}
    \propto\sum_{rsm}
        \uv{e}^\ast\cdot\V{d}_{mr}\Gr_{rs}
        \uv{e}\cdot\V{d}_{sm}.
\end{equation}
Because the sums in $r$ and $s$ go over all excited states, and
$m$ runs over all ground states, we can write
Eq.~\eqref{eq:fluorescencegeneral} for our case in terms of the
uncoupled-basis states. This gives
\begin{widetext}
\begin{equation}
    \mc{I}
    \propto\sum(-1)^{q'+q''}\prn{e^\ast}_{q'}e_{q''}
    \bra{Im_IJ_gm_J}d_{-q'}\ket{Im'_IJ_em'_J}
    \bra{Im'_IJ_em'_J}\Gr\ket{Im''_IJ_em''_J}
    \bra{Im''_IJ_em''_J}d_{-q''}\ket{Im_IJ_gm_J},
\end{equation}
\end{widetext}
resulting in the restrictions
\begin{equation}\label{eq:fluorconditions}
\begin{gathered}
    m''_J=-q''+m_J,\
    m_J=-q'+m'_J,\\
    m_I=m'_I=m''_I,
\end{gathered}
\end{equation}
on the terms that can contribute to the fluorescence. This
indicates that the nuclear polarization cannot affect the
fluorescence signal, and so only the electronic excited-state
polarization of rank $\Gk\le2J_e$ can be observed. In addition,
only coherences with $\abs{m''_J-m'_J}=\abs{q'+q''}\le2$ can be
observed. This rule has appeared earlier as a consequence of
the low-light-power assumption; because spontaneous decay is
not induced by an incident light field, in this case the rule
is exact. This means that no matter the value of $J_e$, and
what polarization moments exist in the excited state, only
polarization of rank $\Gk\le2$ can be observed in fluorescence.

\subsection{Summary}

In this section, we have shown that, when the ground- or
excited-state hfs is unresolved, there are restrictions on the
rank of the polarization moments that can be created or
detected by light. Some of these restrictions may at first seem
counter-intuitive, but they can be obtained from very basic
considerations. For example, the two facts that nuclear spin
can be ignored if the hfs is completely unresolved and that
lowest-order depopulation pumping of a given hyperfine level
does not depend on ground-state hyperfine splitting lead
directly to the result that polarization moments produced by
depopulation pumping are subject to a limit of $\Gk\le2J_g$
when the excited-state hfs is unresolved. Various processes of
creation and detection of polarization are subject to different
restrictions (Table~\ref{tab:summary}).
\begin{table}
\begin{tabular}{c@{\quad}c@{\quad}c}
\hline
    & unresolved & limit on $\Gk$ \\
    \hline
Ground-state pol.\ (depop.) & excited hfs & $\le2J_g$\\
Ground-state pol.\ (repop.) & ground hfs & $\le2J_e$\\
Excited-state pol.\ & ground hfs & $\le2J_e$\\
Absorption & excited hfs & $\le2J_g$\\
Fluorescence & --- & $\le2J_e$\\
\hline
\end{tabular}
\caption{Summary of the results of this section. For each
quantity, the restriction on the rank $\Gk$ of the polarization
that can be created or detected is given in the third column.
The restriction holds under conditions (ground- or
excited-state hfs unresolved) described in the second column.
For fluorescence with broad-band detection the restriction
holds regardless of whether the hfs is resolved.}
\label{tab:summary}
\end{table}
In particular, the two processes that can create ground-state
polarization---depopulation and repopulation pumping---are
subject to restrictions under different conditions.
Consequently, unless the hfs is entirely unresolved, there is
always a mechanism for producing polarization limited in rank
only by the total angular momentum, rather than the electronic
angular momentum.

\section{Partially resolved hyperfine structure: nonlinear magneto-optical effects}
\label{sec:nmoe}

Now let us examine the more general case of \emph{partially}
resolved hyperfine transitions. For this study, we will look at
the quantitative dependence on hyperfine splitting of nonlinear
optical rotation---rotation of light polarization due to
interaction with a $J_g\rightarrow J_e$ transition group in the
presence of a magnetic field. In this case, the effect of
ground-state atomic polarization is brought into starker
relief: in the experimental situation that we consider, both
the creation and detection of ground-state polarization is
required in order to see any signal whatsoever. When linearly
polarized light is used, as is supposed here, the lowest-order
effect depends on rank-two atomic alignment. Thus, for the
alkali atoms, the question of the dependence of the effect on
hyperfine structure arises, because, as discussed in the
previous section, both the creation and the detection of
alignment in the $J_g=1/2$ ground state can be suppressed due
to unresolved hfs. (In fact, a higher-order effect can occur
wherein alignment is created, the alignment is converted to
orientation, and the orientation is detected
\cite{Auz92,Bud2000AOC,Auz2006}. However, the conversion of
alignment to orientation is an effect of tensor ac Stark
shifts, which can be shown by arguments similar to those in
Sec.~\ref{sec:unresolved} to suffer suppression due to
unresolved hfs in the same way as does the direct detection of
alignment.)

\begin{figure}
\includegraphics[width=3.4in]{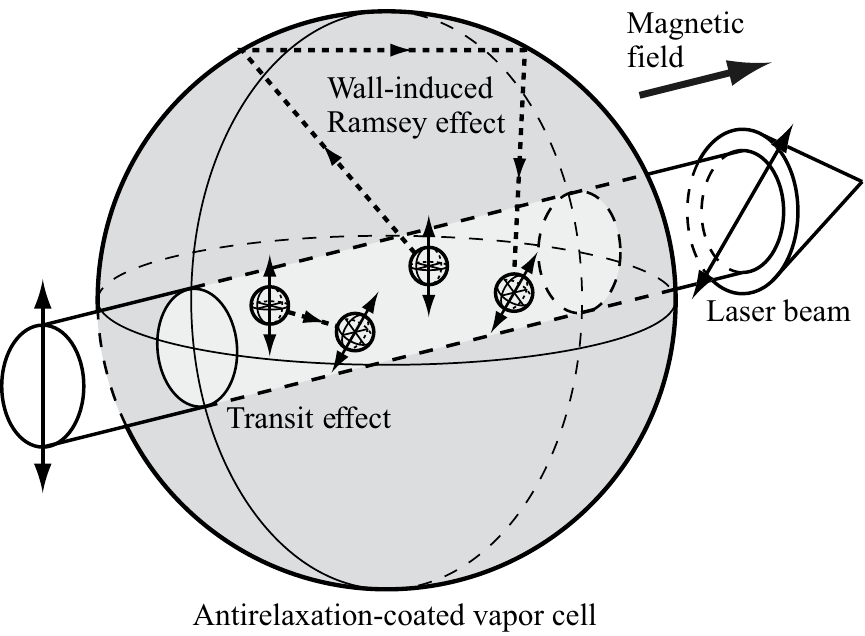}
\caption{The transit and wall optical-rotation effects. Both
effects occur in an anti-relaxation-coated vapor cell, and can
be distinguished by their greatly different magnetic resonance
widths.}
\label{fig:transitandwall}
\end{figure}

In the Faraday geometry, linearly polarized light propagates in
the direction of an applied magnetic field, and the rotation of
the light polarization direction is measured. A number of
magneto-optical effects can contribute to the optical rotation,
including the linear Faraday effect, the Bennett-structure
effect, and various effects depending on atomic polarization
(``coherence effects'') \cite{Bud2002RMP}. Here we are
concerned with optical rotation due to several different forms
of the ground-state coherence effect, in which the atomic
velocities are treated in three different ways. First we
consider the atoms to have no velocity spread, and analyze the
Doppler-free ``transit effect'', as for an atomic beam with
negligible transverse velocity distribution \cite{Sch93}. We
then consider the case in which atoms have a Maxwellian
distribution, but do not change their velocities in between
pumping and probing---this corresponds to the transit effect
for buffer-gas-free, dilute atomic vapors \cite{Kan93}.
Finally, we treat the case in which atoms undergo
velocity-changing collisions between pumping and probing, as
for buffer-gas cells \cite{NovAc2001} or the wall-induced
Ramsey effect (``wall effect'') in antirelaxation-coated vapor
cells \cite{Kan95}. Figure~\ref{fig:transitandwall} illustrates
the transit and wall effects in a vapor cell.
We examine the dependence of these effects on the size of the
hyperfine splittings as they vary from much smaller than the
natural width to much greater than the Doppler width.

Throughout this section we consider formulas for the optical
rotation signal valid to lowest order in light power, under the
assumption that the ground-state relaxation rate $\Gg$ is much
smaller than both the excited-state natural width $\GG$ and the
hyperfine splittings. For the Doppler-free case a single
analytic formula can be applied to both resolved and unresolved
hfs (i.e., no assumption need be made about the relative size
of the hyperfine splittings and the natural width). For the
Doppler-broadened cases, analytic results can be obtained in
various limits, which together describe the signal over the
entire range of hyperfine splittings.


We first focus on the simplest case: the $D1$ line
($J_g=J_e=1/2$) for an atom with $I=1/2$. This is a somewhat
special case, because one of the two ground-state hyperfine
levels has $F_g=0$, and consequently can neither support atomic
alignment nor produce optical rotation. We then consider the
differences that arise when considering higher nuclear spin and
also the $D2$ line ($J_g=1/2$ and $J_e=3/2$). Finally, results
for the ``real-world'' alkali atoms commonly used in
experiments are presented. Some details of the calculation and
general formulas for arbitrary $J_g$, $J_e$, and $I$ are
presented in Appendix~\ref{app:NMORcalc}. These formulas are
generalizations of those given in Ref.~\cite{Kan93}; related
earlier work includes that of Refs.~\cite{Hap67b,Mat70}.

\subsection{Doppler-free transit effect}
\label{subsec:dfeffect}

We consider nonlinear Faraday rotation on a $J_g\rightarrow
J_e$ atomic transition for an atom with nuclear spin $I$. We
can limit our attention to the ground-state coherence effects
by using a ``three-stage'' model for Faraday rotation
\cite{Kan93}, in which optical pumping, atomic precession, and
optical probing take place sequentially, and the light and
magnetic fields are never present at the same time. In this
case, the linear and Bennett-structure effects, which require
the simultaneous application of light and magnetic fields, do
not occur. Such a model can be realized in an atomic beam
experiment, but it is also a good approximation to a vapor cell
experiment that uses low light power and small enough magnetic
fields so that the coherence effects are dominant.

The calculation is performed using second order perturbation
theory in the basis of the polarization moments $\Gr^{(\Gk
q)}(F_1F_2)$ of the density matrix (Appendix~\ref{app:PMPT}).
The three stages of the calculation are as follows. In stage
(a), a $x$-directed light beam linearly polarized along $z$ is
applied, and we calculate optical pumping through second order
in the optical Rabi frequency. In stage (b), the light field is
removed, and a $x$-directed magnetic field is applied. We
calculate the effect of this field on the atomic polarization.
Finally, in stage (c), the magnetic field is turned off, and
the light field is applied once more to probe the atomic
polarization. The nonlinear optical rotation is found to lowest
order in the probe-light Rabi frequency
(Appendix~\ref{app:dfeffect}).

Because the magnetic field is neglected during the optical
pumping stage, the atomic ground-state polarization that is
produced in this stage is entirely along the light polarization
direction, i.e., it has polarization component $q=0$. Since
linearly polarized light has a preferred axis, but no preferred
direction, it can not, in the absence of other fields, produce
atomic polarization with a preferred direction, i.e.,
polarization with odd rank $\Gk$. Also, we have seen in
Sec.~\ref{sec:unresolved} that, to lowest order in the light
power, optical pumping cannot produce polarization moments with
$\Gk>2$. Thus the only ground-state polarization moment with
rank greater than zero that is produced at lowest order has
$\Gk=2$ and $q=0$. We first consider the $D1$ line
($J_g=J_e=1/2$) for an atom with $I=1/2$. In this case, the
only ground-state hyperfine level that can support the
$\Gr^{(20)}(F_gF_g)$ moment has $F_g=1$. (Due to the assumption
that the hyperfine splittings are much greater than the
ground-state relaxation rate, we can ignore ground-state
hyperfine coherences throughout the discussion.) From
Eq.~\eqref{eq:rho20}, the value of this moment is found to be
\begin{equation}\label{eq:rho20D1I12}
\begin{split}
    \Gr^{(20)}(11)
    &=\frac{\Gk_s}{12\sqrt{6}}
        \Big(
        \sbr{L(\Go'_{0,1})-L(\Go'_{1,1})}\\
        &\qquad\qquad\qquad
        +\frac{R}{3}
        \sbr{L(\Go'_{1,0})-L(\Go'_{1,1})}
        \Big),
\end{split}
\end{equation}
where $\Gk_s=\rme{J_g}{d}{J_e}^2\mc{E}_0^2/(\GG\Gg)$ is the
reduced optical-pumping saturation parameter ($\mc{E}_0$ is the
optical electric field amplitude), $R$ is the branching ratio
for the transition $J_e\rightarrow J_g$, and $\Go'_{F_eF_g}$ is
the transition frequency between excited-state and ground-state
hyperfine levels in the frame ``rotating'' at the
Doppler-shifted light frequency $\Go$:
$\Go'_{F_eF_g}=\Go_{F_eF_g}-\Go+\V{k}\cdot\V{v}$, where
$\Go_{F_eF_g}$ is the transition frequency in the lab frame,
$\Go$ is the light frequency, $\V{k}$ is the wave vector, and
$\V{v}$ is the atomic velocity. We also write
$\Go'_{F_eF_g}=-\GD_{F_eF_g}+\V{k}\cdot\V{v}$, where
$\GD_{F_eF_g}$ is the light detuning from resonance. We have
defined the Lorentzian line profile
\begin{equation}\label{eq:lorentzian}
    L(\Go')=\frac{(\GG/2)^2}{(\GG/2)^2+\Go'^2}.
\end{equation}

Equation \eqref{eq:rho20D1I12} is written as the sum of two
terms, each surrounded by square brackets. The first term is
the contribution to the polarization due to depopulation
pumping discussed in Sec.~\ref{sec:depop}. This term is itself
a sum of contributions due to pumping on the $F_g=1\rightarrow
F_e=0$ transition and the $F_g=1\rightarrow F_e=1$ transition.
These two contributions are of opposite sign, as illustrated in
Fig.~\ref{leveldiagD1I12_1to0_1to1}. Pumping on either
transition produces alignment in the $F_g=1$ ground state; the
sign of the corresponding polarization moment depends on
whether there is more population in the $m=0$ sublevel or the
$m=\pm1$ sublevels. We saw in the discussion of
Sec.~\ref{sec:depop} that when the excited state hfs is
unresolved, polarization with rank $\Gk>2J_g$ cannot be created
by depopulation pumping (Fig.~\ref{leveldiagD1I12_1to01}). We
see here that as $\Go_{0,1}$ approaches $\Go_{1,1}$, i.e., as
the excited-state hyperfine splitting goes to zero, the
contributions from the two transitions cancel and this term
goes to zero. For the Doppler-broadened atomic ensemble
discussed in Sec.~\ref{sec:unresolved}, the hfs was considered
unresolved when the hyperfine splittings were smaller than the
Doppler width. Since Eq.~\eqref{eq:rho20D1I12} describes a
single velocity group, the relevant width here is the natural
width $\GG$.

The second term of Eq.~\eqref{eq:rho20D1I12} is the
contribution to the ground-state polarization due to
repopulation pumping discussed in Sec.~\ref{sec:sepol}. This
term is also composed of two contributions of opposite sign:
one due to pumping on the $F_g=1\rightarrow F_e=1$ transition
and one due to pumping on the $F_g=0\rightarrow F_e=1$
transition. The two contributions are illustrated in
Figs.~\ref{leveldiagD1I12_1to1_SE} and
\ref{leveldiagD1I12_0to1_SE}, which show the origin of the
opposite signs. In Sec.~\ref{sec:sepol} we found that
depopulation pumping cannot create polarization moments with
rank $\Gk>2J_e$ when the ground-state hfs is unresolved
(Fig.~\ref{leveldiagD1I12_01to1_SE}). We see here that this
term of Eq.~\eqref{eq:rho20D1I12} goes to zero when $\Go_{1,0}$
approaches $\Go_{1,1}$, i.e., as the ground-state hyperfine
splitting goes to zero.

In the second and third stages of the model of the coherence
effect, the ground-state polarization precesses in a magnetic
field and is probed by light with the same polarization as the
pump light considered in the first stage. From Eq.\
\eqref{eq:normalizedrotation} we find that the normalized
optical rotation $d\Gv$ per path length $d\ell$ is proportional
to the polarization produced in the first stage and is given by
\begin{equation}\label{eq:D1I12Rot}
    \ell_0\frac{d\Gv}{d\ell}
    =\frac{1}{4}\sqrt{\frac{3}{2}}
        \sbr{L(\Go'_{0,1})-L(\Go'_{1,1})}
        x_1\Gr^{(20)}(11),
\end{equation}
where
\begin{equation}\label{eq:magreslineshape}
    x_{F_g}
    =\frac{(\Gg/2)\GO_{F_g}}
        {(\Gg/2)^2+\GO_{F_g}^2}
\end{equation}
is the magnetic-resonance line-shape parameter, with
$\GO_{F_g}=g_{F_g}\Gm_B B$ the Larmor frequency for the
ground-state hyperfine level $F_g$ ($g_{F_g}$ is the Land\'e
factor for the ground state $F_g$, and $\Gm_B$ is the Bohr
magneton), and
\begin{equation}\label{eq:dfabslength}
    \ell_0
    =-\prn{\frac{1}{\mathcal{I}}\frac{d\mathcal{I}}{d\ell}}^{-1}
    =\frac{2\Gp}{Rn\Gl^2}
    \frac{(2J_g+1)}{(2J_e+1)}
\end{equation}
is the unsaturated resonant absorption length assuming totally
unresolved hyperfine structure, where $\mathcal{I}$ is the
light intensity, $n$ is the atomic density, and $\Gl$ is the
light wavelength. The branching ratio $R$ enters here because
it factors into the transition strength.

\begin{figure}
   \includegraphics{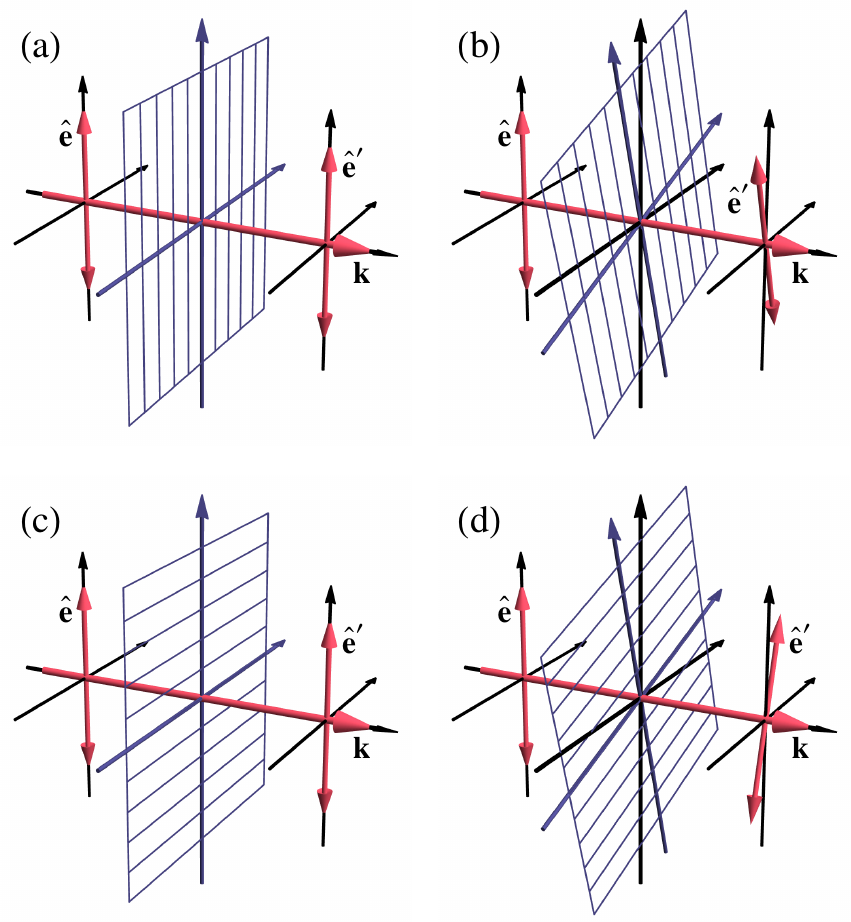}
   \caption{Illustration of the rotating polarizer model for
   optical rotation. (a) Optical pumping on a $F_g=1\rightarrow
   F_e=0$ or $F_g=1\rightarrow F_e=1$ transition causes the
   medium to act as a polarizing filter with transmission axis
   along the input light polarization $\uv{e}$. (b) When the
   transmission axis rotates due to Larmor precession, the
   output light polarization $\uv{e}'$ follows the transmission
   axis and so rotates in the same sense as the Larmor
   precession. (c) If polarization produced by pumping on one
   transition is probed on the other, the polarization
   functions as a polarizing filter with transmission axis
   \emph{perpendicular} to the input light polarization.
   (Attenuation of the light beam is not indicated.) (d) When
   the medium polarization rotates, the output light
   polarization tends to rotate towards the transmission axis,
   in the opposite sense to the Larmor precession in this
   case.} \label{fig:RPM}
\end{figure}

The contributions to the optical rotation signal from the
$F_g=1\rightarrow F_e=0$ transition and the $F_g=1\rightarrow
F_e=1$ transition have opposite signs. To understand this, it
is helpful to think of the optically polarized medium as a
polarizing filter \cite{Kan93}. When pumping on a
$1\rightarrow0$ or $1\rightarrow1$ transition, the medium is
pumped into a dark (non-absorbing) state for that transition
(Fig.~\ref{leveldiagD1I12_1to0_1to1}), corresponding to a
polarizing filter with its transmission axis along the input
light polarization axis $\uv{e}$ [Fig.~\ref{fig:RPM}(a)].
The Larmor precession induced by the magnetic field causes the
transmission axis of the filter to rotate, so that it is no
longer along $\uv{e}$. This in turn causes the output light
polarization axis $\uv{e}'$ to rotate. The polarization of
light passing through a polarizing filter tends to rotate
towards the transmission axis, so that in this case the optical
rotation is in the same sense as the Larmor precession
[Fig.~\ref{fig:RPM}(b)]. Now, compare the polarization produced
when pumping on a $1\rightarrow0$ or $1\rightarrow1$
transition, as shown in Fig.~\ref{leveldiagD1I12_1to0_1to1}. We
see that the dark state for each transition is a bright
(absorbing) state for the other. This means that if we choose
one or the other of these states, it will function as just
described for one of the transitions, but will function as a
polarizing filter with its transmission axis
\emph{perpendicular} to $\uv{e}$ for the other transition
[Fig.~\ref{fig:RPM}(c)]. When the axis of the filter rotates in
this case, the fact that the output light polarization tends to
rotate towards the transmission axis, means that here the
optical rotation is in the other direction, in the opposite
sense to the Larmor precession [Fig.~\ref{fig:RPM}(d)]. In
other words, for a particular sign of the rank-two polarization
moment, the optical rotation will have one sign when probed on
one transition, and the opposite sign when probed on the other,
as indicated by Eq.~\eqref{eq:D1I12Rot}. Because the
observation of optical rotation requires the detection of
rank-two polarization moments, we might expect, analogously to
the discussion in Sec.\ \ref{sec:abs}, that it is suppressed
when the excited-state hyperfine splitting goes to zero.
Equation~\eqref{eq:D1I12Rot} shows that the two contributions
indeed cancel when $\Go_{0,1}$ approaches $\Go_{1,1}$.

\begin{figure}
   \includegraphics{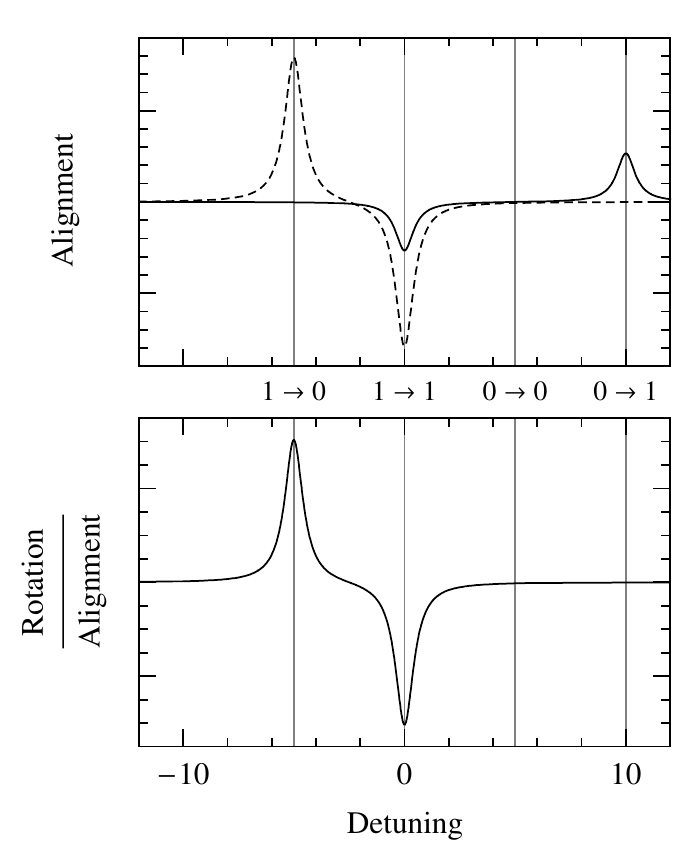}
   \caption{Dependence on light detuning from the
   $F_g=1\rightarrow F_e=1$ transition of (top) the components
   of ground-state alignment due to depopulation (dashed) and
   repopulation (solid) [Eq.~\eqref{eq:rho20D1I12}] and
   (bottom) optical rotation for a given amount of alignment
   [Eq.~\eqref{eq:D1I12Rot}]. Gray vertical lines show $F_g\rightarrow
   F_e$ transition resonance frequencies. Parameter values in units of $\GG$ are
   $\Gg\ll1$, $A_g=10$, $A_e=5$.}
   \label{fig:D1I12PolRot}
\end{figure}

Equation \eqref{eq:D1I12Rot} and the two components of
Eq.~\eqref{eq:rho20D1I12} are plotted as a function of light
detuning from the $F_g=1\rightarrow F_e=1$ transition in
Fig.~\ref{fig:D1I12PolRot}, for particular values of the
ground- and excited-state hyperfine coefficients $A_g$ and
$A_e$.
(For $J=I=1/2$, the hyperfine coefficient $A$ is equal to the
splitting between the two hyperfine levels.) Here and below
numerical values of frequencies are given in units of $\GG$. As
discussed above, each spectrum consists of two peaks of equal
magnitude and opposite sign. For the spectrum of alignment due
to depopulation and the spectrum of rotation for a given amount
of alignment, the peaks are separated by the excited-state
hyperfine splitting, so that they cancel as this splitting goes
to zero. For the spectrum of alignment due to repopulation, the
peaks are separated by the ground-state hyperfine splitting;
they cancel as the ground-state splitting goes to zero.

\begin{figure}
   \includegraphics{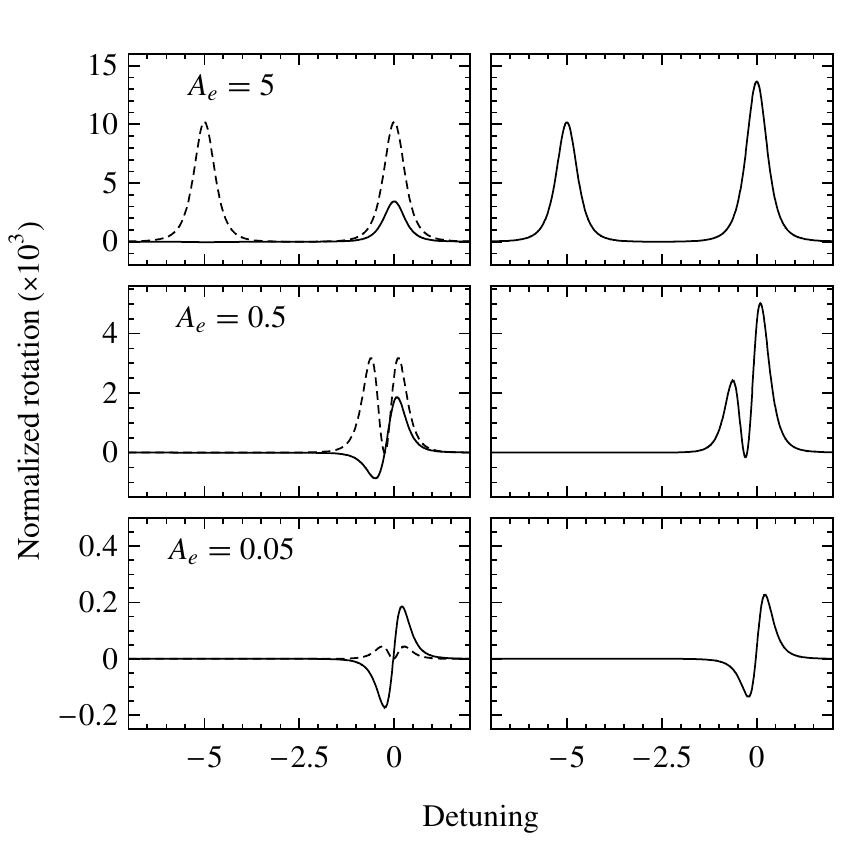}
   \caption{Spectra of normalized optical rotation
   $\ell_0/(\Gk_sx_1)(d\Gv/d\ell)$ for the Doppler-free transit
   effect. Left column: components due to polarization produced
   by depopulation (dashed) and repopulation (solid); right
   column: total signal. Parameter values in units of $\GG$ are
   $\Gg\ll1$, $A_g\gg1,A_e$.} \label{fig:D1I12RotSpectra}
\end{figure}

\begin{figure}
   \includegraphics{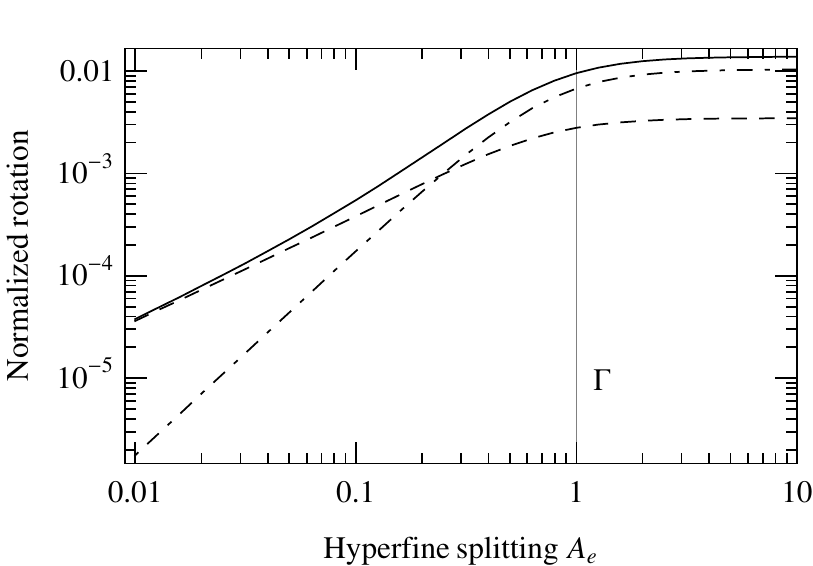}
   \caption{Maximum of the spectrum of the Doppler-free
   nonlinear magneto-optical rotation transit effect as a function of
   excited-state hyperfine splitting. Plotted are the component
   due to polarization produced by depopulation (dash-dotted)
   due to polarization produced by repopulation (dashed) and
   the total signal (solid). Parameters as in
   Fig.~\ref{fig:D1I12RotSpectra}.}
   \label{fig:D1I12MaxRotLargeAg}
\end{figure}

In this subsection we are analyzing a Doppler-free system,
i.e., we assume that the atoms all have the same velocity,
which we take to be zero for simplicity. Then the observed
optical rotation signal is found by simply substituting
Eq.~\eqref{eq:rho20D1I12} into Eq.~\eqref{eq:D1I12Rot}. We
first consider the case in which the ground-state hfs is well
resolved. The rotation signal is plotted in
Fig.~\ref{fig:D1I12RotSpectra} for large ground-state hyperfine
splitting and various excited-state splittings $A_e$.
The components of the rotation signal due to depopulation
(dashed) and repopulation (solid) are plotted in the left-hand
column, and the total signal is plotted on the right. As the
previous discussion indicates, the rotation signal decreases as
the excited-state hyperfine splitting $A_e$ becomes smaller,
with the component due to depopulation decreasing faster than
the component due to repopulation. This is also seen in
Fig.~\ref{fig:D1I12MaxRotLargeAg}, which shows the maximum
magnitude of the rotation spectrum as a function of $A_e$ (for
each value of $A_e$, the signal is optimized with respect to
detuning).
Thus, for small splittings, the component due to repopulation
dominates. To lowest order in $A_e$, the signal is given by
\begin{equation}\label{eq:D1I12DFlargeAgsmallAe}
    \ell_0\frac{d\Gv}{d\ell}
    =\frac{A_e\Gk_s x_1R(\GG/2)^4\GD_{1}}
        {144\sbr{(\GG/2)^2+\GD_{1}^2}^3},
\end{equation}
i.e., linear in $A_e$, with a modified dispersive shape that
falls off far from resonance as $1/\GD_{1}^5$, where $\GD_1$ is
the detuning from the center of the $F_g=1\rightarrow F_e$
transition group.

The previous discussion also explains why the two peaks in the
component due to depopulation seem to cancel as they overlap,
even though they have the same sign: the factors in the signal
due to the creation and detection of alignment cancel
individually (Fig.~\ref{fig:D1I12PolRot}); it is only in their
product that the two peaks have the same sign.

\begin{figure}
   \includegraphics{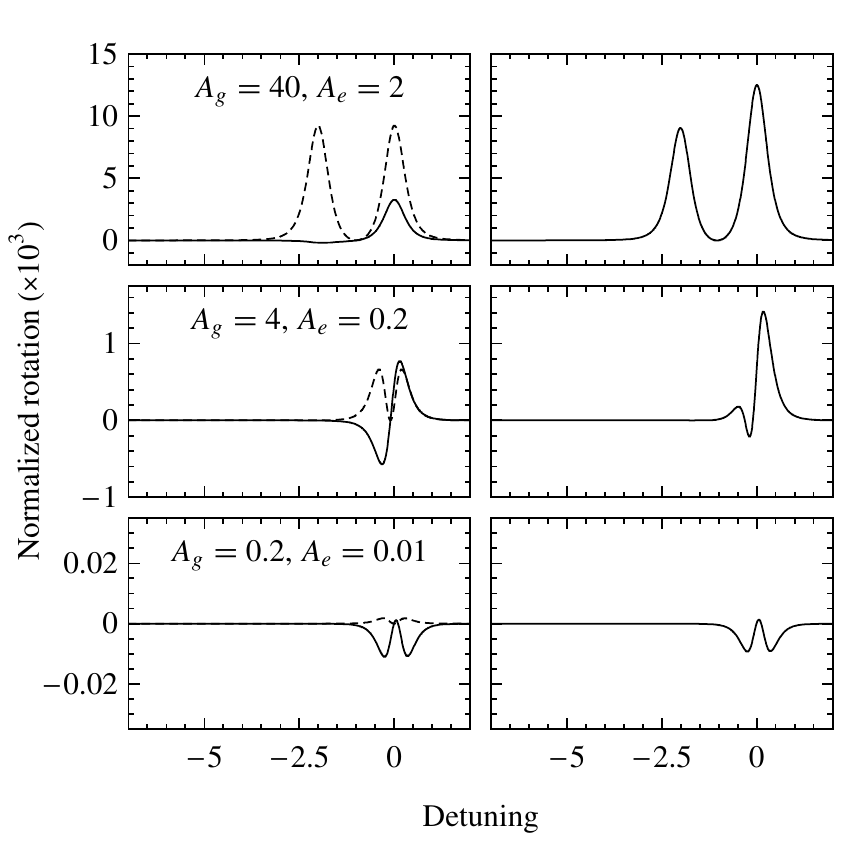}
   \caption{As Fig.~\ref{fig:D1I12RotSpectra}, with $A_g$ and $A_e$
   varied simultaneously ($A_g=20A_e$). } \label{fig:D1I12DFSpectraSmallAgSmallAe}
\end{figure}

We now consider the case in which both the ground- and
excited-state hyperfine splittings are small, so that all of
the hfs is unresolved. To lowest order in $A_g$ and $A_e$ we
have
\begin{equation}\label{eq:D1I12DFsmallAgsmallAe}
    \ell_0\frac{d\Gv}{d\ell}
    =A_e\prn{A_e-\frac{R}{3}A_g}
    \frac{\Gk_sx_1(\GG/2)^4\GD^2}
        {24\sbr{(\GG/2)^2+\GD^2}^4},
\end{equation}
where $\GD$ is the light detuning from the line center of the
$D1$ transition. As we expect, the component of the signal due
to polarization produced by repopulation is proportional to
$A_g$ for small hyperfine splitting. The component of the
signal resulting from depopulation-induced polarization also
enters at this order. The optical rotation spectrum in this
case is double-peaked, and falls off as $1/\GD^6$
(Fig.~\ref{fig:D1I12DFSpectraSmallAgSmallAe}).

\subsection{Doppler-broadened transit effect}
\label{subsec:DBeffect}

We now consider an atomic ensemble with a Maxwellian velocity
distribution, but a low rate of velocity-changing collisions,
so that the atomic velocities do not change between optical
pumping and probing. This is the case for an atomic beam
experiment, or for the ``transit effect'' in a dilute-vapor
cell. Because the atoms have a fixed velocity, the signal from
each velocity group can be found individually and then summed
to find the total signal. Thus the signal from the
Doppler-broadened transit effect is found by multiplying the
Doppler-free signal found in the previous subsection by a
Gaussian weighting function representing the Doppler
distribution along the light propagation direction and then
integrating over atomic velocity. We can perform this integral
analytically in different limiting cases.

We first consider the commonly encountered experimental case in
which the hyperfine splitting is much greater then the natural
line width of the excited state, i.e., the Doppler-free
spectrum is well resolved. In this case, for a given light
frequency and atomic velocity, the light acts on at most one
transition between hyperfine levels. Thus the excited-state
hyperfine coherences can be neglected, and the cancelation
effects due to the overlap of resonance lines do not appear. As
found in Eq.~\eqref{eq:DFrotationresolved}, the Doppler-free
rotation spectrum then appears as a collection of peaks, one
centered at each optical resonance frequency, each with
line-shape function $f(\Go_{F_eF_g}')=L(\Go_{F_eF_g}')^2$,
i.e., the square of a Lorentzian line shape. (One Lorentzian
factor is due to optical pumping, the other to probing.)

In this case, the Doppler-broadened signal is found by making
the replacement $f\rightarrow f_{DB}$, where the velocity
integral for $f_{DB}$ takes the form
\begin{equation}
    f_{DB}(\GD_{F_eF_g})=\int dv_k f(-\GD_{F_eF_g}+k_Bv_k)G(v_k),
\end{equation}
where
\begin{equation}\label{eq:DopplerDist}
    G(v_k)=\frac{k_B}{\GG_D\sqrt{\Gp}}e^{-(k_Bv_k/\GG_D)^2}
\end{equation}
is the normalized distribution of atomic velocities along the
light propagation direction $\uv{k}$, $k_B$ is the Boltzmann
constant, and $\GG_D$ is the Doppler width. This integral can
be evaluated in terms of the error function. Under the
assumption $\GG\ll\GG_D$ that we will employ here, the integral
can be approximated by replacing $f$ with a properly normalized
delta function, resulting in
\begin{equation}
    f_{DB}(\GD_{F_eF_g})
    \approx\frac{\sqrt{\Gp}}{4}\frac{\GG}{\GG_D}e^{-(\GD_{F_eF_g}/\GG_D)^2}.
\end{equation}
The Doppler-broadened spectrum, given explicitly by
Eq.~\eqref{eq:DBrotation}, thus consists of a collection of
resonances, each with Gaussian line shape. For the $D1$ line
with $I=1/2$, we have
\begin{equation}
    \ell_0\frac{d\Gv}{d\ell}
    =\frac{\Gk_sx_1}{576}
        \prn{(3+R)e^{-(\GD_{1,1}/\GG_D)^2}
        +3e^{-(\GD_{0,1}/\GG_D)^2}}.
    \label{eq:largelowerdopplerlargeA}
\end{equation}
Here $\ell_0$ is the absorption length for the
Doppler-broadened case, given by
\begin{equation}\label{eq:dbabslength}
    \ell_0
    =\frac{4\sqrt{\Gp}}{Rn\Gl^2}
    \frac{\GG_D}{\GG}
    \frac{(2J_g+1)}{(2J_e+1)}.
\end{equation}
Equation \eqref{eq:largelowerdopplerlargeA} is valid for
$A_e,A_g,\GG_D\gg\GG$. Note that all the terms in this
expression have the same sign; thus no cancelation occurs when
the resonances overlap. This is because the Doppler-free
resonances all have the same sign when the Doppler-free
spectrum is well resolved (Fig.~\ref{fig:D1I12RotSpectra}), so
when the Doppler-broadened spectrum samples more than one
resonance, the contributions from each resonance add.

The same approach can be generalized to describe the case in
which some or all of the hyperfine splittings are on the order
of or smaller than $\GG$. In this case, the Doppler-free
spectrum is not composed entirely of peaks with a shape given
by $f(\Go_{F_eF_g}')$. Nevertheless, as long as each resonance
or group of resonances has frequency extent much less than the
Doppler width, we can approximate it as a delta function times
a coefficient given by the integral of the Doppler-free
spectrum over the resonance. For the $D1$ line with $I=1/2$ and
$A_e,\GG\ll\GG_D\ll A_g$, this procedure yields
[Eq.~\eqref{eq:DBrotsmallAe}]
\begin{equation}
    \ell_0\frac{d\Gv}{d\ell}
    =\frac{A_e^2\Gk_sx_1(6+R)e^{-\GD_{1,1}^2/\GG_D^2}}
    {576\prn{\GG^2+A_e^2}}.
    \label{eq:largelowerdopplersmallA}
\end{equation}
The rotation in this case goes as $A_e^2$ for small $A_e$; the
term linear in $A_e$ [Eq.~\eqref{eq:D1I12DFlargeAgsmallAe}] is
odd in detuning and consequently cancels in the velocity
integral.

\begin{figure}
    \includegraphics{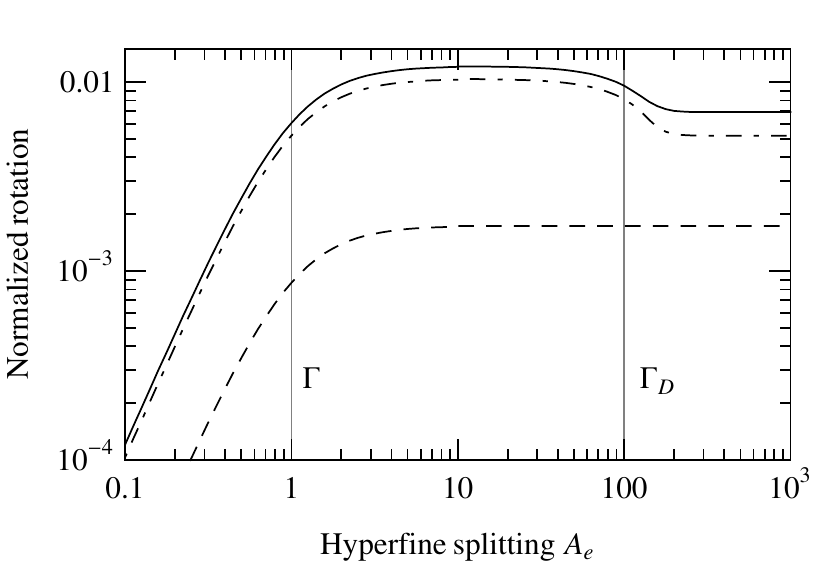}
    \caption{Maximum of the spectrum of the Doppler-broadened
    effect as a function of excited-state hyperfine splitting.
    Plotted are the component due to polarization produced by
    depopulation (dash-dotted), due to polarization produced by
    repopulation (dashed), and the total signal (solid).
    Equation \eqref{eq:largelowerdopplersmallA} is used for
    $A_e<10$, and Eq.~\eqref{eq:largelowerdopplerlargeA} is
    used for $A_e>10$. Parameter values in units of $\GG$ are $\GG_D=100$,
    $\Gg\ll1$, $A_g\gg\GG_D$.}
    \label{fig:D1I12MaxRotLargeAgDB}
\end{figure}

Since Eq.~\eqref{eq:largelowerdopplerlargeA} applies when
$A_e\gg\GG$ and Eq.~\eqref{eq:largelowerdopplersmallA} applies
when $A_e\ll\GG_D$, we have that---if $\GG_D$ is sufficiently
larger than $\GG$---the two formulas together describe the
signal over the entire range of $A_e$ to excellent
approximation, as verified by a numerical calculation.
Figure~\ref{fig:D1I12MaxRotLargeAgDB} shows the maximum of the
rotation spectrum as a function of the excited-state hyperfine
splitting.
As discussed above, as $A_e$ is reduced, there is no
suppression of the optical rotation signal when the
Doppler-broadened hfs becomes unresolved. Only when the
Doppler-free spectrum for a particular velocity group becomes
unresolved is there suppression, as described in the previous
subsection.

\begin{figure}
   \includegraphics{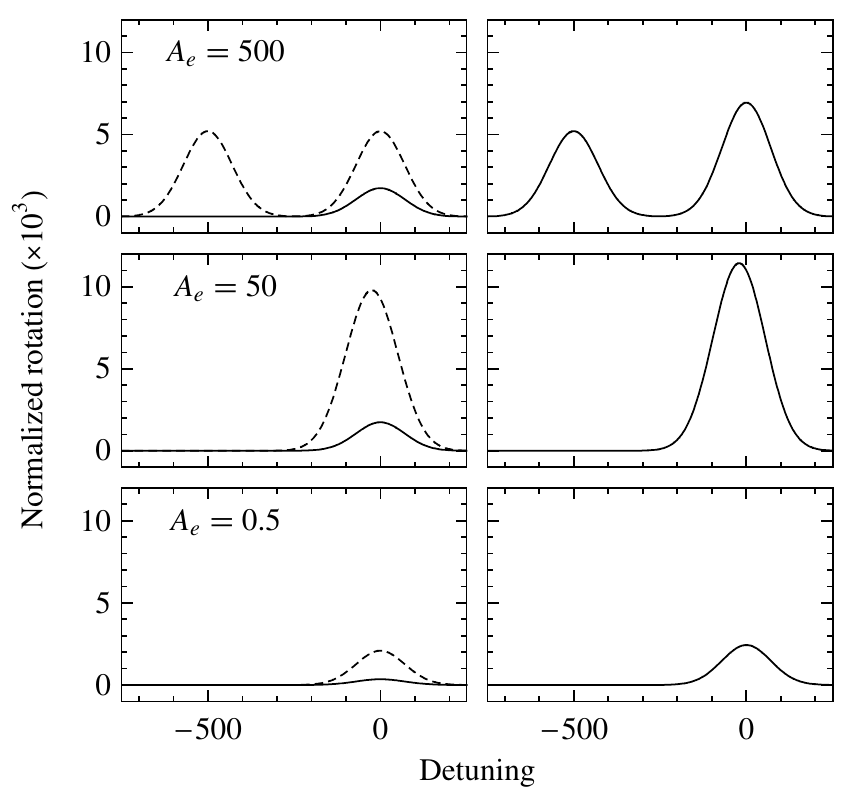}
   \caption{Spectra, as Fig.~\ref{fig:D1I12RotSpectra}, but for the
   Doppler-broadened transit effect. Parameter values in units of $\GG$ are
   $\GG_D=100$, $\Gg\ll1$, $A_g\gg\GG_D$.}
   \label{D1I12RotSpectraDBLargeAg}
\end{figure}

\begin{figure}
   \includegraphics{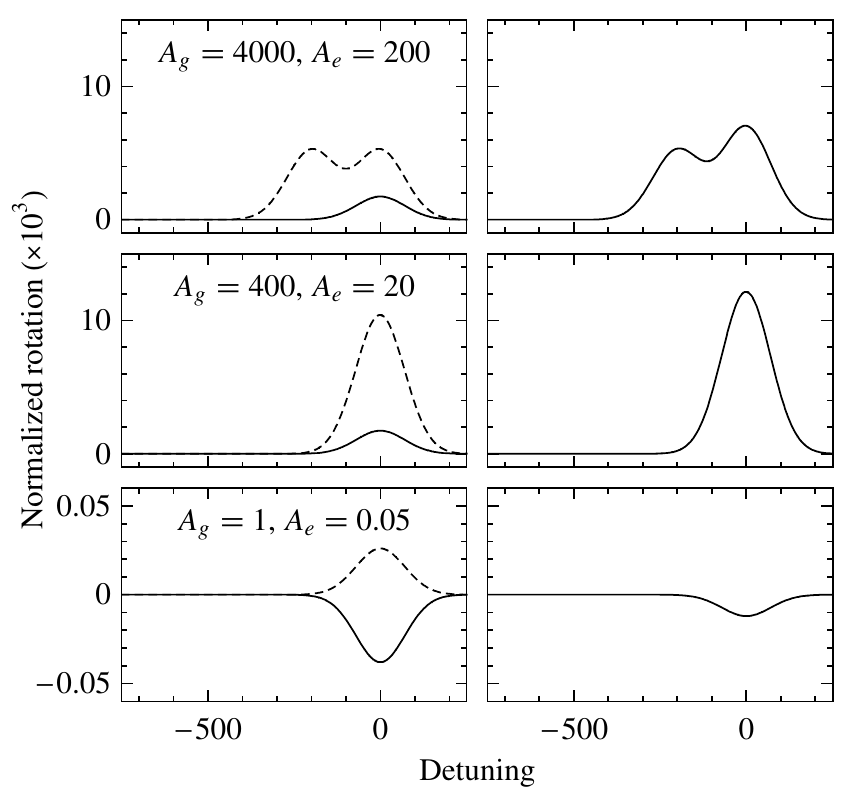}
   \caption{As Fig.~\ref{D1I12RotSpectraDBLargeAg}, with $A_g$ and
   $A_e$ varied simultaneously ($A_g=20A_e$).} \label{D1I12RotSpectraDBEqualAgAe}
\end{figure}

Spectra for the Doppler-broadened transit effect are shown in
Fig.~\ref{D1I12RotSpectraDBLargeAg} for large $A_g$ and various
values of $A_e$, and for $A_g$ and $A_e$ varied together in
Fig.~\ref{D1I12RotSpectraDBEqualAgAe}.

In the case in which both $A_e$ and $A_g$ are small, the
Doppler-free rotation spectrum is entirely of the same sign
[see Eq.~\eqref{eq:D1I12DFsmallAgsmallAe} and the bottom plot
of Fig.~\ref{fig:D1I12DFSpectraSmallAgSmallAe}]. The
Doppler-broadened signal thus behaves similarly to the
Doppler-free signal, because no additional cancelation takes
place upon integrating over the velocity distribution. The
signal for the $D1$ line with $I=1/2$ and $A_g,A_e,\GG\ll\GG_D$
is given by
\begin{equation}\label{eq:D1I12DBsmallAgsmallAe}
    \ell_0\frac{d\Gv}{d\ell}
    =A_e\prn{A_e-\frac{R}{3}A_g}
    \frac{\Gk_sx_1}{96\,\GG^2}
    e^{-\GD^2/\GG_D^2}.
\end{equation}

\subsection{Wall effect}
\label{subsec:walleffect}

We now consider systems in which the atomic velocities change
in between optical pumping and probing. This is the case for
the ``wall effect'' in antirelaxation-coated vapor cells: atoms
are optically pumped as they pass through the light beam, and
then retain their polarization through many collisions with the
cell walls before returning to the beam and being probed
(Fig.~\ref{fig:transitandwall}). A similar situation occurs in
vapor cells with buffer gas.

We assume that the atomic velocities are completely randomized
after optical pumping. Then the density matrix for each
velocity group is the same; to lowest order in light power, we
can find the velocity-averaged polarization by integrating the
perturbative expression \eqref{eq:rho20} over velocity with the
Gaussian weighting factor \eqref{eq:DopplerDist}. Since we are
now describing the average over all of the atoms in the cell,
and not just the illuminated region of the cell, we take $\Gg$
to be the average ground-state relaxation rate for an atom in
the cell, rather than the transit rate through the light beam.
We also multiply the polarization by the illuminated fraction
of the cell volume, $V_\text{illum.}/V_\text{cell}$ (assuming
this fraction is small), to account for the fact that the light
pumps only some of the atoms at a time.

For the specific case of the $D1$ line for an atom with
$I=1/2$, Eq.~\eqref{eq:rho20} takes the form, given in Eq.\
\eqref{eq:rho20D1I12}, of a linear combination of Lorentzian
functions $L(\Go_{F_eF_g}')$. This simple form arises because,
due to the selection rules for this transition, no coherences
are formed between excited-state hyperfine levels. For a
general system this is not the case; however, if the
excited-state hyperfine splitting is greater than $\GG$, the
excited-state hyperfine coherences are suppressed, and all
resonances once again have Lorentzian line shapes. Thus,
assuming that $\GG\ll\GG_D$, the velocity integral can be
accomplished by replacing $L(\Go_{F_eF_g}')$ by
\begin{equation}
    \int dv_k L(-\GD_{F_eF_g}+kv_k)G(v_k)
    \approx \frac{\sqrt{\Gp}}{2}
    \frac{\GG}{\GG_D}e^{-(\GD_{F_eF_g}/\GG_D)^2}.
\end{equation}
The polarization in this case is given by
[Eq.~\eqref{eq:Wallpol}]
\begin{equation}\label{eq:rho20D1I12Wall}
\begin{split}
    \Gr^{(20)}(11)
    &=\frac{\Gk_s\sqrt{\Gp}}{24\sqrt{6}}
        \Big(
        \sbr{e^{-(\GD_{0,1}/\GG_D)^2}-e^{-(\GD_{1,1}/\GG_D)^2}}\\
        &\qquad\qquad
        +\frac{R}{3}
        \sbr{e^{-(\GD_{1,0}/\GG_D)^2}-e^{-(\GD_{1,1}/\GG_D)^2}}
        \Big),
\end{split}
\end{equation}
where the saturation parameter for the wall effect is defined
by
\begin{equation}\label{eq:wallsatparam}
    \Gk_s=\frac{\GO_R^2}{\GG\Gg}\frac{\GG}{\GG_D}
    \frac{V_\text{illum.}}{V_\text{cell}}.
\end{equation}
We make this new definition because, in the wall effect, light
of a single frequency illuminating just part of the cell
effectively pumps all velocity groups in the entire cell.

\begin{figure}
   \includegraphics{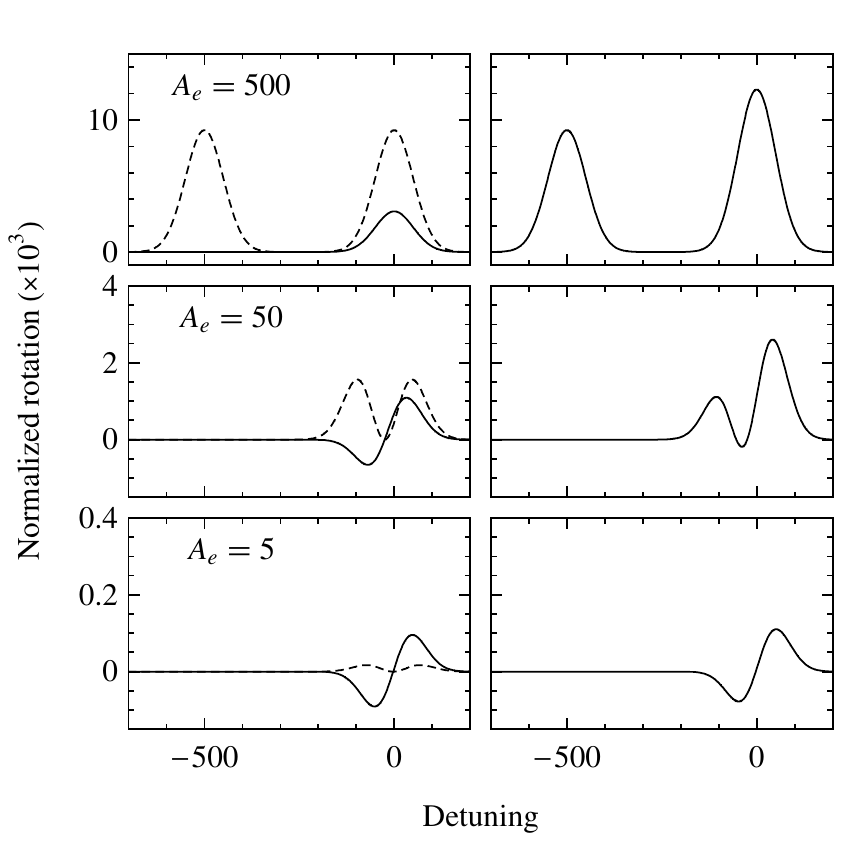}
   \caption{Spectra, as Fig.~\ref{D1I12RotSpectraDBLargeAg}, but for
   the wall effect.}
   \label{fig:D1I12RotSpectraWall}
\end{figure}

\begin{figure}
    \includegraphics{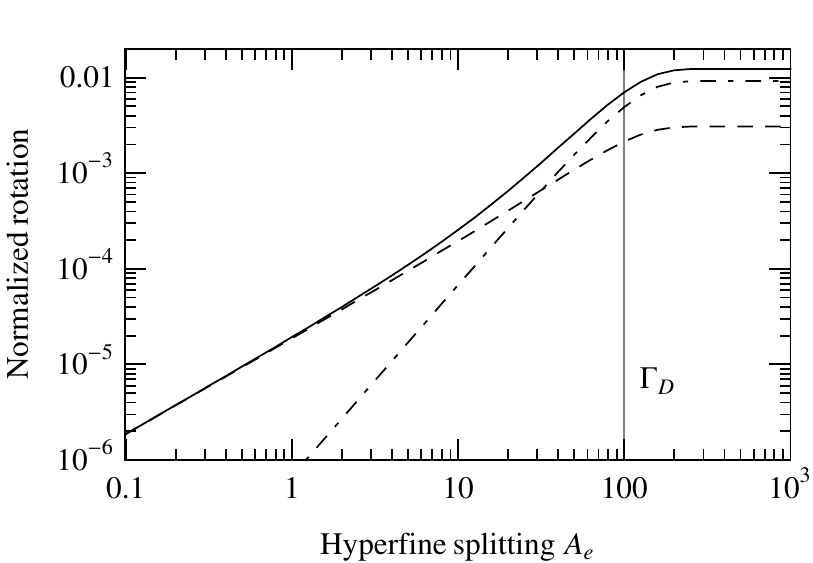}
    \caption{As Fig.~\ref{fig:D1I12MaxRotLargeAg}, but for the wall
    effect. Parameters are the same as in Fig.\
    \ref{fig:D1I12MaxRotLargeAgDB}.}
    \label{fig:D1I12MaxRotLargeAgWall}
\end{figure}

\begin{figure}
   \includegraphics{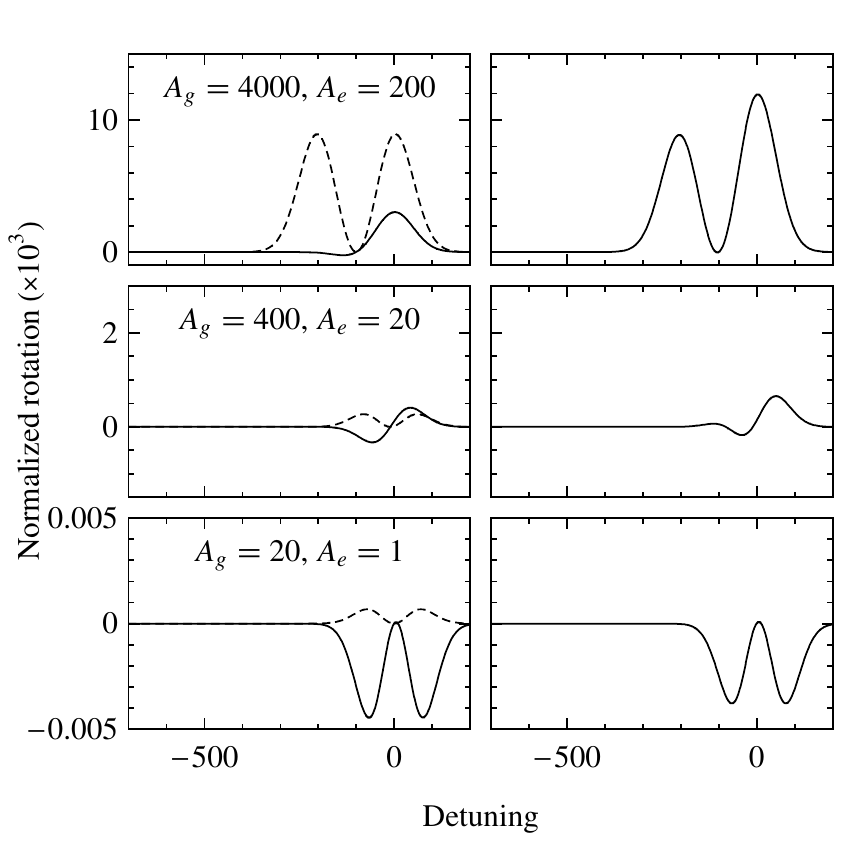}
   \caption{As Fig.~\ref{D1I12RotSpectraDBEqualAgAe}, but for the wall
   effect.} \label{fig:D1I12SpectraSmallAgSmallAeWall}
\end{figure}

The signal due to each velocity group is given in terms of
$\Gr^{(20)}(1)$ by Eq.~\eqref{eq:D1I12Rot}; integrating over
velocity to find the total signal, we obtain
[Eq.~\eqref{eq:WallRotation}]
\begin{equation}\label{eq:D1I12RotWall}
    \ell_0\frac{d\Gv}{d\ell}
    =\frac{1}{4}\sqrt{\frac{3}{2}}
        \sbr{e^{-(\GD_{0,1}/\GG_D)^2}-e^{-(\GD_{1,1}/\GG_D)^2}}
        x_1\Gr^{(20)}(11).
\end{equation}
The spectrum of the signal due to the wall effect is quite
different than the spectrum of the Doppler-broadened transit
effect signal, and is in a sense more similar to that of the
Doppler-free transit effect \cite{Bud2000Sens}. Equations
\eqref{eq:rho20D1I12Wall} and \eqref{eq:D1I12RotWall} have the
same form as the Doppler-free Eqs.~\eqref{eq:rho20D1I12} and
\eqref{eq:D1I12Rot}, with Lorentzians of width $\GG$ replaced
by Gaussians of width $\GG_D$. Thus the rotation signal
produced by the wall effect has similar spectra and dependence
on hyperfine splitting as the Doppler-free transit effect, but
with scale set by the Doppler width rather than the natural
width. This is illustrated in
Figs.~\ref{fig:D1I12RotSpectraWall} and
\ref{fig:D1I12MaxRotLargeAgWall} for the case of large
ground-state hyperfine splitting.
Figure \ref{fig:D1I12RotSpectraWall} shows the optical rotation
spectrum for various values of $A_e$, and
Fig.~\ref{fig:D1I12MaxRotLargeAgWall} shows the maximum of the
rotation spectrum as a function of $A_e$. These figures can be
compared to Figs.~\ref{fig:D1I12RotSpectra} and
\ref{fig:D1I12MaxRotLargeAg} for the Doppler-free transit
effect. In particular, we see the same phenomenon of two
resonance peaks of the same sign appearing to cancel as they
overlap (observation of this effect in anti-relaxation-coated
vapor cells is discussed in Ref.~\cite{Bud2000Sens} and in
buffer-gas cells in Ref.~\cite{NovAc2001}). The explanation for
this is the same as in the Doppler-free case. Also as in the
Doppler-free case, the rotation is linear in $A_e$ to lowest
order, and this linear term is due to polarization produced by
spontaneous emission:
\begin{equation}\label{eq:D1I12WallLargeAgsmallAe}
    \ell_0\frac{d\Gv}{d\ell}
    =\frac{\sqrt{\Gp}}{288}
    \Gk_sx_1R
    \frac{A_e\GD_{1}}{\GG_D^2}
    e^{-2(\GD_{1}/\GG_D)^2}.
\end{equation}
Spectra for the case in which $A_e$ and $A_g$ are varied
together are shown in
Fig.~\ref{fig:D1I12SpectraSmallAgSmallAeWall}, and are also
similar to the Doppler-free transit effect
(Fig.~\ref{fig:D1I12DFSpectraSmallAgSmallAe}).
When both $A_e$ and $A_g$ are small, the signal to lowest order
in these quantities is given by
\begin{equation}\label{eq:D1I12WallSmallAgsmallAe}
    \ell_0\frac{d\Gv}{d\ell}
    =A_e\prn{A_e-\frac{R}{3}A_g}
    \frac{\Gk_sx_1\GD^2}{48\,\GG_D^4}
    e^{-2\GD^2/\GG_D^2}.
\end{equation}

\subsection{Higher nuclear spin and the $D2$ line}

When nuclear spins $I\ge1/2$ are considered, several
complications arise. The clearest of these is that the two
ground states now have angular momenta
$F_g=I\pm\frac{1}{2}\ge1$, so that they can both support atomic
alignment and produce optical rotation. A more subtle
difference is that, with higher angular momenta in the excited
state, coherences between the excited state hyperfine levels
can be created when the excited-state hyperfine splitting is on
the order of the natural width or smaller. (Ground-state
hyperfine coherences can be neglected as long as the
ground-state hyperfine splitting is much larger than the
ground-state relaxation rate.) This can change the optical
rotation spectrum, and also causes the symmetry between the
Doppler-free transit and wall effects discussed above to be
partially broken, as we see below.

However, many of the results obtained above for the $I=1/2$
system are a consequence of the general arguments discussed in
Sec.~\ref{sec:unresolved}, and thus hold for any nuclear spin.
In particular, the dependence of the optical rotation signal on
the hyperfine splitting for large ground-state and small
excited state splitting [Eqs.\
\eqref{eq:D1I12DFlargeAgsmallAe},
\eqref{eq:largelowerdopplersmallA}, and
\eqref{eq:D1I12WallLargeAgsmallAe}] and for both ground- and
excited-state hyperfine splitting small [Eqs.\
\eqref{eq:D1I12DFsmallAgsmallAe},
\eqref{eq:D1I12DBsmallAgsmallAe}, and
\eqref{eq:D1I12WallSmallAgsmallAe}] remains the same. We have,
for large $A_g$ and small $A_e$, and for a particular
transition group, the following three expressions. For the
Doppler-free transit effect:
\begin{equation}\label{eq:D1DFLargeAgSmallAe}
    \quad\ell_0\frac{d\Gv}{d\ell}
    \propto A_e\Gk_s x_{F_g}R
        \frac{(\GG/2)^4\GD_{F_g}}
        {\big[(\GG/2)^2+\GD_{F_g}^2\big]^3},
\end{equation}
for the Doppler-broadened transit effect:
\begin{equation}\label{eq:D1DBLargeAgSmallAe}
    \ell_0\frac{d\Gv}{d\ell}
    \propto A_e^2\Gk_sx_{F_g}
    \frac{e^{-\GD_{F_g}^2/\GG_D^2}}
    {\prn{\GG^2+A_e^2}},
\end{equation}
and for the wall effect:
\begin{equation}\label{eq:D1WallLargeAgSmallAe}
    \ell_0\frac{d\Gv}{d\ell}
    \propto A_e\Gk_sx_{F_g}R
    \frac{\GD_{F_g}e^{-2(\GD_{F_g}/\GG_D)^2}}{\GG_D^2}.
\end{equation}
For  $A_g$ and $A_e$ both small, we have, for the Doppler-free
transit effect:
\begin{equation}\label{eq:D1DFSmallAgSmallAe}
    \ell_0\frac{d\Gv}{d\ell}
    \propto A_e\prn{A_e-\frac{R}{3}A_g}
    \frac{\Gk_s(\GG/2)^4\GD^2}
        {\sbr{(\GG/2)^2+\GD^2}^4},
\end{equation}
for the Doppler-broadened effect:
\begin{equation}\label{eq:D1DBSmallAgSmallAe}
    \ell_0\frac{d\Gv}{d\ell}
    \propto A_e\prn{A_e-\frac{R}{3}A_g}
    \frac{\Gk_s}{\GG^2}
    e^{-\GD^2/\GG_D^2},
\end{equation}
and for the wall effect:
\begin{equation}\label{eq:D1WallSmallAgSmallAe}
    \ell_0\frac{d\Gv}{d\ell}
    \propto A_e\prn{A_e-\frac{R}{3}A_g}
    \frac{\Gk_s\GD^2}{\GG_D^4}
    e^{-2\GD^2/\GG_D^2}.
\end{equation}

\begin{figure*}
   \includegraphics{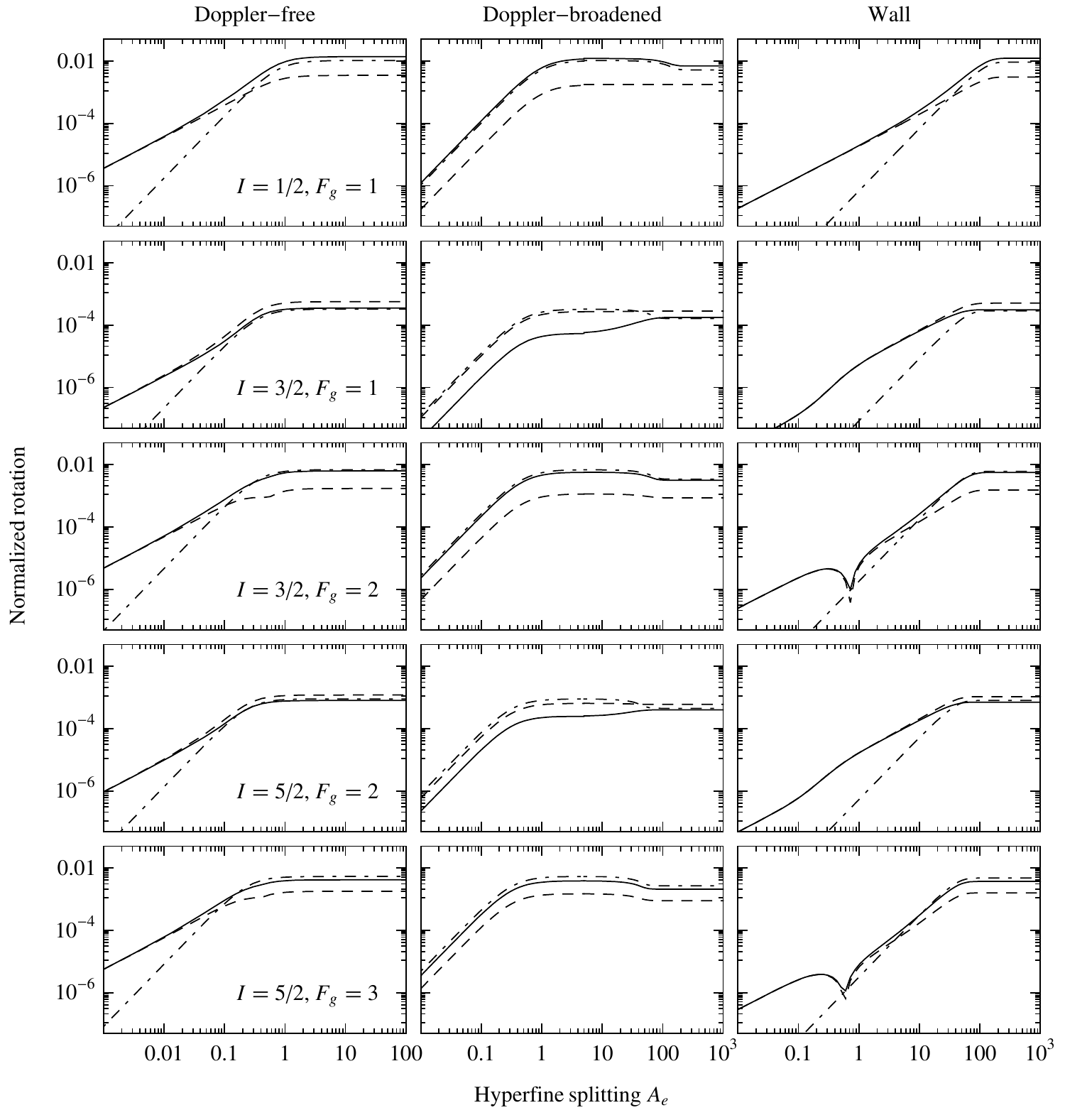}
   \caption{Maximum of the normalized optical rotation spectra
   $\ell_0/(\Gk_sx_{F_g})d\Gv/d\ell$ for the Doppler-free
   transit, Doppler-broadened transit, and wall effects on the
   $D1$ line for $I=1/2$, 3/2, and 5/2. We assume $\GG_D=100$,
   $\Gg\ll1$, $A_g\gg\GG_D$ in units of $\GG$. The maxima for
   the $F_g=I\pm1/2\rightarrow F_e$ transitions are plotted
   separately. Each plot shows rotation due to polarization
   produced by depopulation (dot-dashed line), rotation due to
   polarization produced by repopulation (dashed line), and
   total rotation (solid line). The sharp features seen in some
   of the plots occur when two terms contributing to the largest
   resonance cancel. In general, since all resonances are not
   canceled at the same time, the maximum of the spectrum does
   not go to zero.} \label{fig:D1LargeAgTable}
\end{figure*}

To illustrate the differences that arise when the nuclear spin
is increased, we plot (analogously to
Figs.~\ref{fig:D1I12MaxRotLargeAg},
\ref{fig:D1I12MaxRotLargeAgDB}, and
\ref{fig:D1I12MaxRotLargeAgWall}) in
Fig.~\ref{fig:D1LargeAgTable} the maximum of the rotation
spectra for large $A_g$ as a function of $A_e$ for the
Doppler-free transit, Doppler-broadened transit, and wall
effects. Three values of the nuclear spin are used, $I=1/2$,
3/2, and 5/2, and for $I=3/2$ and 5/2 the rotation on the
$F_g=I\pm1/2\rightarrow F_e$ lines is plotted separately.
Rotation due to polarization produced by the depopulation and
repopulation mechanisms is plotted, as well as the total
rotation signal. In many cases these two contributions are of
opposite sign, so the details of the total signal can depend on
how closely the two contributions cancel each other. (The
cancelation tends to be more complete for the $F_g=I-1/2$
lines.) However, the qualitative features of these plots
follow, in large part, the pattern exhibited in the $I=1/2$
case. One exception is the behavior of the wall effect plot for
$A_e$ in the neighborhood of the natural width. As mentioned
above, when $I>1/2$, excited-state hyperfine coherences can
form when the excited-state hyperfine splitting becomes small.
This leads to ``interference'' effects when the Doppler-free
resonance lines overlap that do not occur when the
Doppler-broadened resonance lines in the wall effect overlap.
This breaks the symmetry between the wall effect and the
Doppler-free transit effect that is found in the $I=1/2$ case.

\begin{figure*}
   \includegraphics{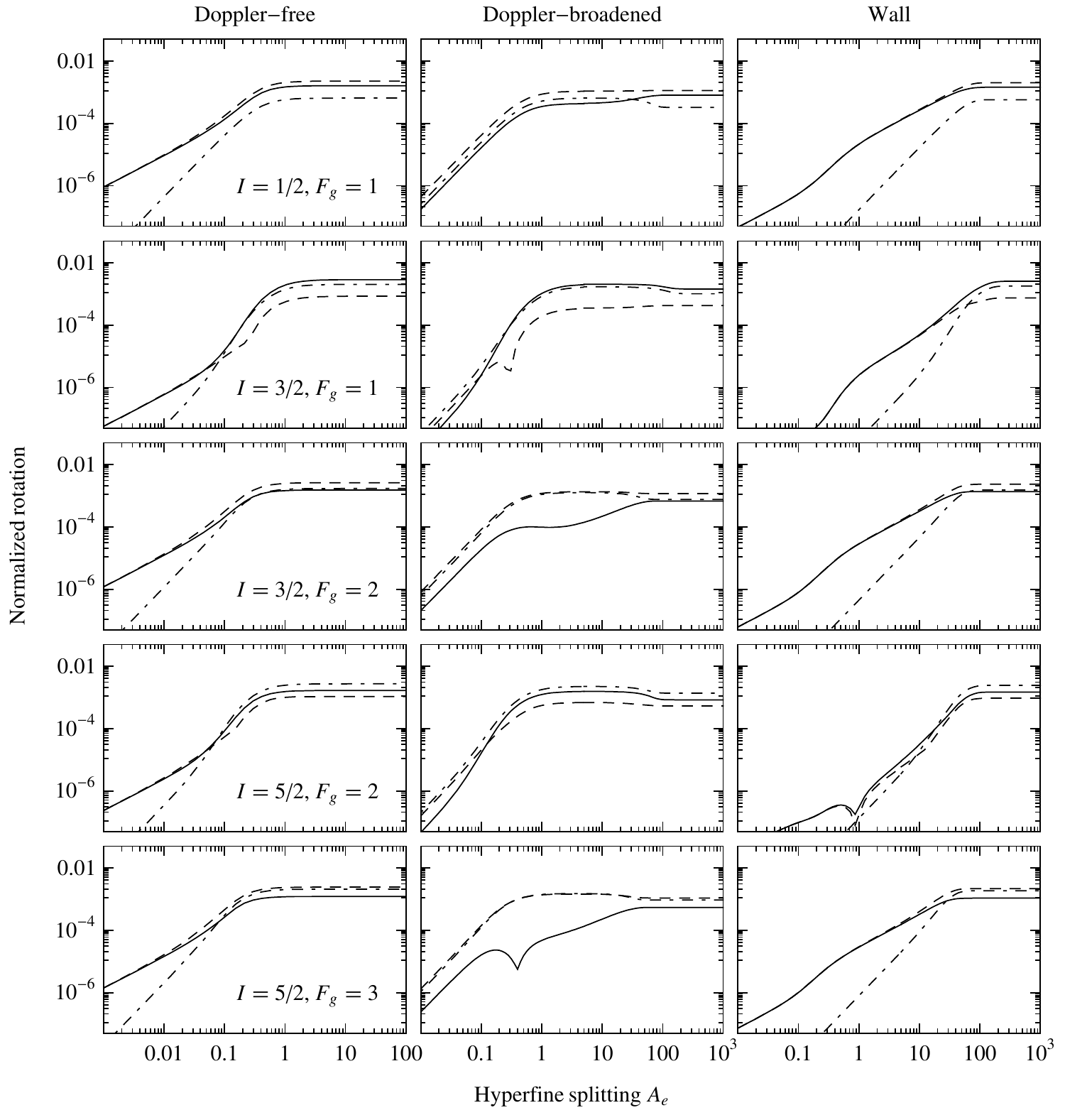}
   \caption{Maximum of the normalized optical rotation spectra,
   as in Fig.~\ref{fig:D1LargeAgTable}, but for the $D2$ line.
   For the $I=3/2$, $F_g=2$ and the $I=5/2$, $F_g=3$ systems
   for the Doppler-broadened transit effect, the two
   contributions to optical rotation nearly cancel, with the
   consequence that the approximations used in obtaining the
   analytic formulas for the total Doppler-broadened signal
   begin to break down. Numerical convolution is employed in
   these cases.} \label{fig:D2LargeAgTable}
\end{figure*}

We now discuss the $J_g=1/2\rightarrow J_e=3/2$ $D2$
transition. The presence of three hyperfine levels in the
excited state leads to additional features in the dependence of
the signal on the hyperfine splitting
(Fig.~\ref{fig:D2LargeAgTable}).
However, the fact that the ground-state electronic momentum is
still $J_g=1/2$ means that the dependence of the signal on the
excited-state hyperfine splitting as $A_e$ goes to zero remains
the same, for the reasons discussed in
Sec.~\ref{sec:unresolved}. Thus, to lowest order in $A_e$, the
rotation signals on the $D2$ line for large $A_g$ are given by
Eqs.~(\ref{eq:D1DFLargeAgSmallAe}--\ref{eq:D1WallLargeAgSmallAe}).
(We set the hyperfine coefficient $B_e$ to zero for
simplicity.)

Considering the signals obtained when both the excited- and
ground-state hyperfine splittings are small, we expect somewhat
different behavior for the contribution due to polarization
produced by repopulation pumping than in the $D1$ case. This is
because the excited-state electronic angular momentum is
$J_e=3/2$, so that production of rank $\Gk=2<2J_e$ atomic
alignment in the ground-state by spontaneous emission is
allowed even when the ground-state hfs is unresolved
(Sec.~\ref{sec:sepol}). The lowest order dependence on
hyperfine splitting for the $D2$ line is given by
\begin{equation}
    \ell_0\frac{d\Gv}{d\ell}
    \propto A_e\sbr{A_e-R\prn{2A_e+\frac{1}{3}A_g}}
\end{equation}
for each of the three effects, with the spectral line shapes
remaining as in
Eqs.~(\ref{eq:D1DFSmallAgSmallAe}--\ref{eq:D1WallSmallAgSmallAe}).
Note that there is now a term that depends on polarization due
to repopulation that does not go to zero as $A_g$ goes to zero.

\subsection{The alkalis}

\begin{figure*}
    \includegraphics{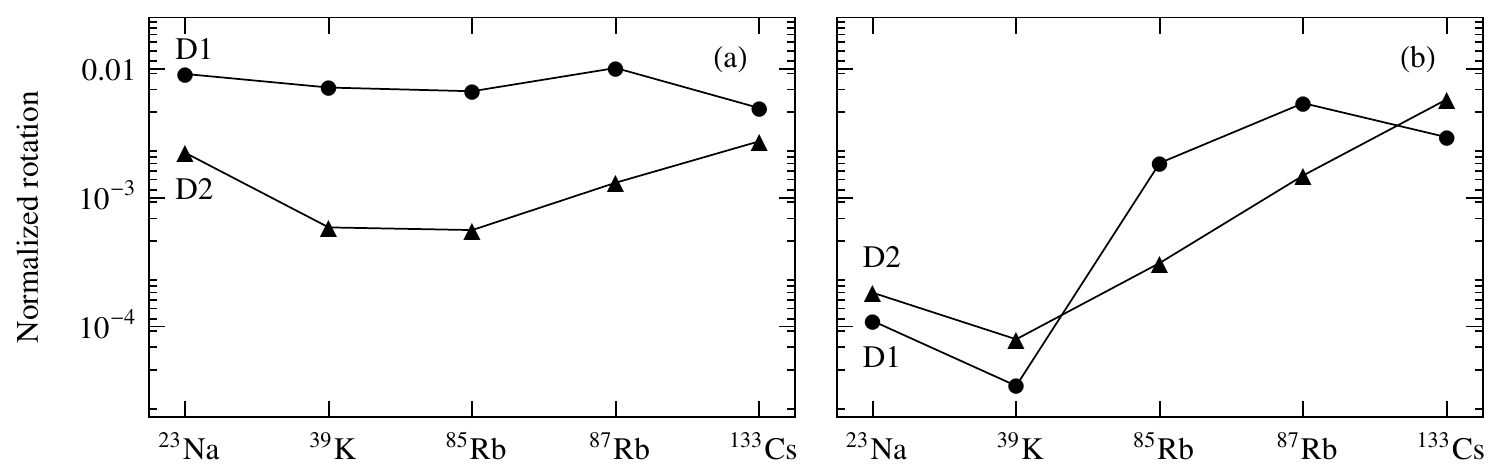}
    \caption{Maximum of the spectrum of normalized optical
    rotation for the (a) Doppler-broadened transit effect and
    (b) wall effect for various alkali atoms. Circles indicate
    the D1 line and triangles indicate the D2 line. Room
    temperature Maxwellian velocity distributions are assumed.
    Normalized rotation is defined here as
    $\ell_0/(\Gk_sx_{I+J_g})(d\Gv/d\ell)$, where $\ell_0$ in
    this case is the unsaturated absorption length at the
    detuning that gives maximum absorption. The normalized
    magnitude of unsuppressed optical rotation is nominally on
    the order of unity; however, this is to some degree
    dependent on the normalization convention chosen. For
    example, if the maximum matrix element of $d_z$ is used in
    the definition of $\Gk_s$, rather than the reduced matrix
    element, the values in this plot are increased by a factor
    of $\sim\!6$.} \label{fig:Alkalis}
\end{figure*}

We now examine the consequences of the preceding discussion for
the alkali atoms commonly used in nonlinear magneto-optical
experiments. In Fig.~\ref{fig:Alkalis} the maximum of the
spectrum of optical rotation is plotted for the D1 and D2 lines
of several alkali atoms.
The Doppler-broadened transit effect is shown in
Fig.~\ref{fig:Alkalis}a and the wall effect is shown in
Fig.~\ref{fig:Alkalis}b. (Numerical convolution was used to
obtain these results, because the alkalis do not all satisfy
the conditions under which the analytic formulas were derived.)
The nuclear spins, hyperfine splittings, excited state
lifetimes, and Doppler widths all vary between the different
alkali atoms. However, focusing our attention on the hyperfine
splittings, which have the greatest degree of variation, we can
see the correspondence of these results to the preceding
discussion. In particular, we have seen that the magnitude of
the Doppler-broadened transit effect is largely independent of
the hyperfine splitting when the splittings are greater than
the natural width of the transition. This is generally the case
for the alkalis, leading to the relative constancy of the
magnitude of the transit effect among the alkalis. For the wall
effect, on the other hand, we have found that the magnitude of
the effect diminishes when the hyperfine splitting becomes less
than the Doppler width. In the alkalis the excited state
hyperfine splitting is generally on the order of or smaller
than the Doppler width, and the general trend is that the ratio
of hyperfine splitting to Doppler width increases as the atomic
mass number increases. This accounts for the general upward
trend in Fig.~\ref{fig:Alkalis}b. The trend is not completely
consistent: the hyperfine splitting of K is smaller than that
of Na, which is reflected in the plot of the wall effect.

\section{Conclusion}

In experiments involving light-induced polarization in the
alkali atoms, the effect of partially resolved hyperfine
structure is of practical importance. We have addressed this
question from both descriptive and quantitative standpoints. We
have formulated rules describing various restrictions on the
rank of atomic polarization moments that can be created or
detected by light in cases when either the ground- or
excited-state hyperfine structure is completely unresolved. We
have also studied the particular situation of nonlinear Faraday
rotation under various experimental conditions in more
generality, and presented analytic formulas giving the results
of optical rotation measurements when the hfs is unresolved,
partially resolved, or completely resolved.

\acknowledgments

The authors acknowledge helpful discussions with
J.~E.~Stalnaker, D.~F.~Jackson Kimball, J.~M.~Higbe, and
W.~Gawlik. This work has been supported by the US ONR MURI and
NGA NURI programs. One of the authors (M.A.) is supported by
Latvian Science Foundation grant 09.1196 and the University of
Latvia grant system; another author (S.M.R.) is supported by a
NASA Earth and Space Science Fellowship.

\appendix

\section{Nonlinear magneto-optical rotation with hyperfine structure}
\label{app:NMORcalc}

\subsection{Perturbation theory with polarization moments}
\label{app:PMPT}

The time evolution of the atomic density matrix $\Gr$ under the
action of a time-independent Hamiltonian $H$ is given by the
Liouville equation, which can be derived from the Schr\"odinger
equation (with phenomenological relaxation terms added by
hand). Setting $\hbar=c=1$, the Liouville equation can be
written
\begin{equation}\label{eq:liouville}
    \dot{\Gr}=-i\sbr{H,\Gr}-\frac{1}{2}\cbr{\GG,\Gr}+\GL+\tr\prn{\mc{F}\Gr},
\end{equation}
where the square brackets denote the commutator and the curly
brackets the anticommutator, $\GG$ is a Hermitian relaxation
matrix that accounts for relaxation mechanisms such as transit
relaxation due to atoms leaving the region of interest and
intrinsic relaxation of excited states due to spontaneous
emission, $\GL$ accounts for repopulation mechanisms, such as
transit repopulation, that do not depend on $\Gr$, and $\mc{F}$
is the spontaneous emission operator, accounting for
repopulation of ground states due to spontaneous emission from
excited states. (We neglect other relaxation and repopulation
mechanisms, such as spin-exchanging collisions, which may
require the inclusion of additional terms). The spontaneous
emission operator, defined by \cite{Bar61a,Bar61b}
\begin{equation}\label{eq:flop}
    \mc{F}^{sr}_{mn}
    =\frac{4}{3}\Go_{rm}^3\V{d}_{mr}\cdot\V{d}_{sn},
\end{equation}
connects a pair of excited states $\ket{s}$, $\ket{r}$ to a
pair of ground states $\ket{m}$, $\ket{n}$; the trace in
Eq.~\ref{eq:liouville} is taken over the excited state pair.

We can expand the operators appearing in the Liouville equation
in terms of the polarization operators $T^{(\Gk)}_q(F_1F_2)$
according to
\begin{equation}
    A=\sum A^{(\Gk q)}(F_1F_2)T^{(\Gk)}_q(F_1F_2),
\end{equation}
where $F_{1,2}$ runs over all pairs of states in the system.
(Here $F$ is understood to represent the total angular momentum
quantum number as well as any additional quantum numbers
necessary to distinguish between two states with the same total
angular momentum.) The expansion coefficients $A^{(\Gk
q)}(F_1F_2)$ can be found from the Wigner-Eckart theorem, along
with the orthonormality condition
\begin{equation}\label{eq:POorthog}
    \tr\prn{
        T^{(\Gk)}_q(F_1F_2)
        \prn{T^{(\Gk')}_{q'}(F_1'F_2')}^\dag}
    =\Gd_{\Gk\Gk'}\Gd_{qq'}
    \Gd_{F_1F_1'}\Gd_{F_2F_2'}
\end{equation}
and the phase convention
\begin{equation}\label{eq:POphase}
\begin{split}
    \prn{T^{(\Gk)}_q(F_1F_2)}^\dag
    &=(-1)^{F_1-F_2+q}T^{(\Gk)}_{-q}(F_2F_1)\\
    &=T^{(\Gk q)}(F_1F_2).
\end{split}
\end{equation}
There are several equivalent expressions for the expansion
coefficients; one such form is
\begin{widetext}
\begin{equation}
\begin{split}
    A^{(\Gk q)}(F_1F_2)
    &=\sum_{mm'}(-1)^{F_1-F_2+q}
        \sqrt{\frac{2\Gk+1}{2F_2+1}}
        \cg{F_1m\Gk,-q}{F_2m'}
        A_{F_1m,F_2m'}.\\
\end{split}
\end{equation}

The set of expansion coefficients for the density matrix are
known as polarization moments. Performing the expansion of each
operator, and using appropriate tensor product and sum rules,
the equation of motion for the polarization moments is found
from the Liouville equation to be
\begin{equation}\label{eq:liouvillePM}
\begin{split}
    \dot{\Gr}^{(\Gk q)}(F_1F_2)
    =&-i(-1)^{F_1+F_2+\Gk}
        \sum
        \sqrt{(2\Gk'+1)(2\Gk''+1)}
        \cg{\Gk'q'\Gk''q''}{\Gk q}
        \sixj(\Gk',\Gk'',\Gk)(F_2,F_1,F_3)\\
        &\qquad\qquad
        \times\Bigg[
        \prn{H^{(\Gk'q')}(F_1F_3)
        -\frac{i}{2}
        \GG^{(\Gk'q')}(F_1F_3)}
        \Gr^{(\Gk''q'')}(F_3F_2)\\
        &\qquad\qquad\qquad\qquad
        -\Gr^{(\Gk'q')}(F_1F_3)
        \prn{H^{(\Gk''q'')}(F_3F_2)
        +\frac{i}{2}
        \GG^{(\Gk''q'')}(F_3F_2)}
        \Bigg]\\
        &+\GL^{(\Gk q)}(F_1F_2)
        +\frac{4}{3}\Go_0^3\sum
            \rme{F_1}{d}{F_e}
            \Gr^{(\Gk q)}(F_eF_e')
            \rme{F_e'}{d}{F_2}
            (-1)^{F_e+F_e'+\Gk+1}
            \sixj(\Gk,F_2,F_1)(1,F_e,F_e'),
\end{split}
\end{equation}
where all variables not appearing on the left-hand side are
summed over (the variables $F_e$ and $F_e'$ appearing in the
last term relate to spontaneous emission and run over only
those states of higher energy than $F_{1,2}$). Here the arrays
enclosed in in curly brackets are the $6j$ symbols.

We now suppose that the total Hamiltonian $H=H_0+V$, where
$H_0$ is diagonal and $V$ is a time-independent perturbation.
We also assume that $\GG$ and $\GL$ are diagonal. More
precisely, we assume that only $\GG^{(00)}(FF)$,
$\GL^{(00)}(FF)$, and $H^{(00)}(FF)$ are nonzero (for arbitrary
$F$). Taking the steady-state limit in
Eq.~\eqref{eq:liouvillePM} and expanding to second order in the
perturbation $V$, we find for a ground-state polarization
moment
\begin{equation}\label{eq:perturbPM}
\begin{split}
    \Gr^{(\Gk q)}(F_gF_g)
    &=\frac{\Gg}{i\ot_{FF}N_g}
        \Bigg[
        \Gd_{\Gk0}\Gd_{q0}\sqrt{2F_g+1}
        -(-1)^{2F_g+\Gk'+\Gk''}
        \sqrt{(2\Gk'+1)(2\Gk''+1)}
        \cg{\Gk'q'\Gk''q''}{\Gk q}\\
        &\qquad\qquad\qquad\times
        \sixj(\Gk'',\Gk',\Gk)(F_g,F_g,F')
        \frac{\ot_{F_gF'}+\ot_{F'F_g}}{\ot_{F_gF_g}\ot_{F_gF'}\ot_{F'F_g}}
        V^{(\Gk'q')}(F'F_g)V^{(\Gk''q'')}(F_gF')\\
        &\qquad\qquad-i\frac{4}{3}\Go_0
        (-1)^{2F'_1+2F'_2+\Gk+\Gk'+\Gk''}
        \sqrt{(2\Gk'+1)(2\Gk''+1)}
        \cg{\Gk'q'\Gk''q''}{\Gk q}
        \sixj(\Gk'',\Gk',\Gk)(F'_2,F'_1,F')
        \sixj(\Gk,F_g,F_g)(1,F'_1,F'_2)\\
        &\qquad\qquad\qquad\times
        \rme{F'_2}{d}{F_g}\rme{F_g}{d}{F'_1}
        \frac{\ot_{F'_1F'}+\ot_{F'F'_2}}
        {\ot_{F'_1F'}\ot_{F'_1F'_2}\ot_{F'F'}\ot_{F'F'_2}}
        V^{(\Gk''q'')}(F'_1F')V^{(\Gk'q')}(F'F'_2)
        \Bigg].
\end{split}
\end{equation}
\end{widetext}
Here we have neglected the possibility of cascade decays and
assumed that $V$ does not couple a state to itself. We also
have assumed that $\GL$ repopulates all ground-state sublevels
equally; i.e. $\GL_{Fm,Fm}=\Gg/N_g$, where $\Gg$ is the
ground-state relaxation rate and $N_g$ is the total number of
ground-state sublevels. The complex energy splitting
$\ot_{F_1F_2}$ is given by
\begin{equation}
\begin{split}
    \ot_{F_1F_2}
    =E_{F_1}-E_{F_2}-\frac{i}{2}\prn{\GG_{F_1}+\GG_{F_2}},
\end{split}
\end{equation}
where $E_F=(2F+1)^{-1/2}H_0^{(00)}(FF)$ is the unperturbed
energy and $\GG_F=(2F+1)^{-1/2}\GG^{(00)}(FF)$ is the total
relaxation rate of a state $F$.

\subsection{Doppler-free transit effect}
\label{app:dfeffect}

We now apply the results obtained in Sec.~\ref{app:PMPT} to the
three-stage calculation described in
Sec.~\ref{subsec:dfeffect}. In stage (a), we consider a
$\uv{z}$-polarized optical electric field
    $\V{\mc{E}}=\mc{E}_0\re\prn{\uv{e}e^{i(\V{k}\cdot\V{r}-\Go t)}}$,
where $\mc{E}_0$ is the optical electric field amplitude,
$\uv{e}=\uv{z}$ is the polarization vector, $\uv{k}=\uv{x}$ is
the wave vector, and $\Go$ is the optical frequency. We let $V$
represent the electric-dipole Hamiltonian in the rotating-wave
approximation: $V'=-\frac{1}{2}d_z \mc{E}_0$. (Here the prime
refers to the rotating frame.) We assume that the magnetic
field is absent in this stage. From Eq.~\eqref{eq:perturbPM} we
find
\begin{widetext}
\begin{equation}\label{eq:rho20}
\begin{split}
    \Gr_a^{(20)}(F_gF_g)
    &=-\sqrt{\frac{2}{3}}
        \sum_{F_e}
        (-1)^{F_g-F_e}\Gk_s
        \frac{(2F_e+1)(2F_g+1)}{(2I+1)(2J_g+1)}
        \Bigg(
        (-1)^{2I+2J_g}
        \sixj(1,1,2)(F_g,F_g,F_e)
        \sixj(J_e,F_e,I)(F_g,J_g,1)^2
        L(\Go'_{F_eF_g})\\
        &\qquad\qquad
        +R\sum_{F_g'F_e'}
        (-1)^{F_g'-F_e'}
        (2J_e+1)
        (2F_g'+1)(2F_e'+1)
        \sixj(1,1,2)(F_e,F_e',F_g')
        \sixj(F_g,F_g,2)(F_e,F_e',1)\\
        &\qquad\qquad\qquad\times
        \sixj(J_e,F_e,I)(F_g,J_g,1)
        \sixj(J_e,F_e,I)(F_g',J_g,1)
        \sixj(J_e,F_e',I)(F_g,J_g,1)
        \sixj(J_e,F_e',I)(F_g',J_g,1)
        \frac{L(\Go'_{F_eF_g'})L(\Go'_{F_e'F_g'})}
             {L\!\prn{\sqrt{\vphantom{y}\smash[b]{\Go'_{F_eF_g'}\Go'_{F_e'F_g'}}}}}
        \Bigg),
\end{split}
\end{equation}
\end{widetext}
where all variables are as defined in
Sec.~\ref{subsec:dfeffect}. We have evaluated matrix elements
using the Wigner-Eckart theorem
and have used the relation (see, for example,
Ref.~\cite{Sob92})
\begin{equation}\label{eq:Gamma0}
    R\,\GG=\frac{4}{3}\frac{\Go^3}{2J_e+1}\rme{J_g}{d}{J_e}^2.
\end{equation}

The unperturbed energies can be evaluated with
\begin{equation}
\begin{split}
    E_{JFM}
    &=E_{J}+\frac{1}{2}K_{IJF}A_{J}\\
    &+\frac{3}{8}
    \frac{K_{IJF}(K_{IJF}+1)-\frac{4}{3}I(I+1)J(J+1)}{I(2I-1)J(2J-1)}
    B_{J},
\end{split}
\end{equation}
where $K_{IJF}=F(F+1)-I(I+1)-J(J+1)$ and $A_{J}$ and $B_{J}$
are the hyperfine coefficients. The last term is zero for
$J\le1/2$ or $I\le1/2$.

In the case in which the excited-state hfs is well resolved in
the Doppler-free spectrum ($\Go_{F_eF_e'}\gg\GG$),
Eq.~\eqref{eq:rho20} reduces to
\begin{widetext}
\begin{equation}
\begin{split}
    &\Gr_a^{(20)}(F_gF_g)
    =-\sqrt{\frac{2}{3}}
        \sum_{F_e}
        (-1)^{F_g-F_e}\Gk_s
        \frac{(2F_e+1)(2F_g+1)}{(2I+1)(2J_g+1)}
        \Bigg(
        (-1)^{2I+2J_g}
        \sixj(1,1,2)(F_g,F_g,F_e)
        \sixj(J_e,F_e,I)(F_g,J_g,1)^2
        L(\Go'_{F_eF_g})\\
        &\quad
        +\sum_{F_g'}
        R\,(-1)^{F_g'-F_e}
        (2J_e+1)
        (2F_g'+1)(2F_e+1)
        \sixj(1,1,2)(F_e,F_e,F_g')
        \sixj(F_g,F_g,2)(F_e,F_e,1)
        \sixj(J_e,F_e,I)(F_g,J_g,1)^2
        \sixj(J_e,F_e,I)(F_g',J_g,1)^2
        L(\Go'_{F_eF_g'})
        \Bigg).\\
\end{split}
\end{equation}
\end{widetext}

In stage (b), the ground-state density matrix, which is
initially in the state found in stage (a), evolves under the
influence of a magnetic field $B\uv{x}$. We will require only
the value of the polarization moment $\Gr_b^{(21)}(F_gF_g)$.
Using the Hamiltonian $H_B=-\V{\Gm}\cdot\V{B}$ in
Eq.~\eqref{eq:liouvillePM} and solving for the steady state, we
find
\begin{equation}\label{eq:stageb}
\begin{split}
   \Gr_b^{(21)}(F_gF_g)
   =i\frac{\sqrt{3}}{2\sqrt{2}}x_{F_g}
   \Gr_a^{(20)}(F_gF_g).
\end{split}
\end{equation}
where the magnetic-resonance line-shape parameter $x_{F_g}$ is
defined in Eq.~\eqref{eq:magreslineshape}.

In stage (c) the ground-state polarization is probed. The
effect of the atoms on the light polarization as the light
traverses the atomic medium can be found in terms of coherences
between ground and excited states using the wave equation:
\begin{equation}\label{eq:waveeq}
    \frac{\partial\V{\mc{E}}}{\partial\ell^2}
    -\frac{\partial\V{\mc{E}}}{\partial t^2}
    =4\Gp\frac{\partial\V{P}}{\partial t^2},
\end{equation}
where $\ell$ is the distance along the light propagation
direction, and $\V{P}=n\tr\Gr\V{d}$ is the medium dipole
polarization ($n$ is atomic density), which can be written in
terms of the rotating-frame density matrix $\Gr'$ as
\begin{equation}\label{eq:Pdef}
\begin{split}
    \V{P}
    =n\sum_{mp}2\re\prn{\Gr'_{pm}\V{d}_{mp}
    e^{i(\V{k}\cdot\V{r}-\Go t)}},
\end{split}
\end{equation}
where $m$ runs over ground states, and $p$ runs over excited
states. Using the parameterization of a general optical
electric field in terms of the polarization angle $\Gv$ and
ellipticity $\Ge$,
\begin{equation}\label{eq:Efromphiepsilon}
\begin{split}
     \V{\mc{E}}
     &=\mc{E}_0\re\Big(e^{i(\V{k}\cdot\V{r}-\Go t+\Gf)}
        [\prn{\cos\Gv\cos\Ge-i\sin\Gv\sin\Ge}\uv{e}_1\\
        &\qquad\qquad{}+\prn{\sin\Gv\cos\Ge+i\cos\Gv\sin\Ge}\uv{e}_2]\Big),
\end{split}
\end{equation}
where $\uv{e}_{1,2}$ are orthogonal transverse unit vectors, we
obtain for optical rotation in the case of linear polarization
\begin{equation}\label{eq:generalOformula}
\begin{split}
    \frac{d\Gv}{d\ell}
    &=-\frac{4\Gp\Go n}{\mc{E}_0}\sum
        \im\sbr{\Gr'_{pm}\V{d}_{mp}\cdot(\uv{k}\times\uv{e})}.
\end{split}
\end{equation}
Using first-order perturbation theory for the optical
coherences and neglecting coherences between nondegenerate
ground states, we obtain the optical rotation for weak probe
light in terms of the ground-state density matrix:
\begin{equation}\label{eq:genrotformula}
    \frac{d\Gv}{d\ell}
    =-2\Gp\Go n
        \im\sbr{\uv{e}\cdot \Gb\cdot(\uv{k}\times\uv{e})},
\end{equation}
where we have defined
\begin{equation}\label{eq:Bdef}
\begin{split}
    \Gb
    &=\sum_{pmn}
        \frac{\V{d}_{pn}
        \Gr_{nm}\V{d}_{mp}}{\ot_{pm}'}.
\end{split}
\end{equation}
Expanding the tensor $\Gb$ in terms of the ground-state
polarization moments, we obtain
\begin{widetext}
\begin{equation}
\begin{split}
    \Gb
    &=\sum_{F_gF_e\Gk q'q''}
        \frac{(-1)^{F_g+F_e+\Gk}}{\ot_{F_eF_g}'}
        \uv{\Ge}_{-q'}\uv{\Ge}_{-q''}
        \cg{1q'1q''}{\Gk,q'+q''}
        \sixj(1,1,\Gk)(F_g,F_g,F_e)
        \abs{\rme{F_g}{d}{F_e}}^2
        \Gr^{(\Gk,q'+q'')}(F_gF_g),
\end{split}
\end{equation}
where $\uv{\Ge}_q$ are the spherical basis vectors.

Evaluating \eqref{eq:genrotformula} for our case and using
Eq.~\eqref{eq:stageb} gives
\begin{equation}\label{eq:normalizedrotation}
\begin{split}
    \ell_0\frac{d\Gv}{d\ell}
    &=-\frac{3\sqrt{3}}{4\sqrt{2}}
        \sum_{F_gF_e}
        (-1)^{F_g+F_e}
        (2F_g+1)(2F_e+1)(2J_g+1)
        \sixj(1,1,2)(F_g,F_g,F_e)
        \sixj(J_e,F_e,I)(F_g,J_g,1)^2
        L(\Go_{F_eF_g}')
        x_{F_g}\Gr_a^{(20)}(F_gF_g),
\end{split}
\end{equation}
where the unsaturated absorption length for the $J_g\rightarrow
J_e$ transition is defined in Eq.~\eqref{eq:dfabslength}.
Substituting in Eq.~\eqref{eq:rho20} results in the complete
expression for optical rotation for the Doppler-free transit
effect:
\begin{equation}
\begin{split}
    \ell_0\frac{d\Gv}{d\ell}
    &=\frac{3}{4}\Gk_s
        \sum_{F_gF_eF_e''}
        (-1)^{2F_g+F_e''-F_e}
        \frac{(2F_e+1)(2F_e''+1)(2F_g+1)^2}{(2I+1)}
        \sixj(1,1,2)(F_g,F_g,F_e'')
        \sixj(J_e,F_e'',I)(F_g,J_g,1)^2
        x_{F_g}\\
    &\qquad\times
        \Bigg(
        (-1)^{2I+2J_g}
        \sixj(1,1,2)(F_g,F_g,F_e)
        \sixj(J_e,F_e,I)(F_g,J_g,1)^2
        L(\Go'_{F_eF_g})
        L(\Go_{F_e''F_g}')\\
        &\qquad\qquad
        +R\sum_{F_g'F_e'}
        (-1)^{F_g'-F_e'}
        (2J_e+1)
        (2F_g'+1)(2F_e'+1)
        \sixj(1,1,2)(F_e,F_e',F_g')
        \sixj(F_g,F_g,2)(F_e,F_e',1)\\
        &\qquad\qquad\qquad\times
        \sixj(J_e,F_e,I)(F_g,J_g,1)
        \sixj(J_e,F_e,I)(F_g',J_g,1)
        \sixj(J_e,F_e',I)(F_g,J_g,1)
        \sixj(J_e,F_e',I)(F_g',J_g,1)
        \frac{L(\Go'_{F_eF_g'})L(\Go'_{F_e'F_g'})L(\Go_{F_e''F_g}')}
             {L\!\prn{\sqrt{\vphantom{y}\smash[b]{\Go'_{F_eF_g'}\Go'_{F_e'F_g'}}}}}
        \Bigg).
\end{split}
\end{equation}


For completely resolved hfs
($\Go_{F_eF_e'},\Go_{F_gF_g'}\gg\GG$), this reduces to
\begin{equation}\label{eq:DFrotationresolved}
\begin{split}
    \ell_0\frac{d\Gv}{d\ell}
    &=\frac{3}{4}\Gk_s
        \sum_{F_gF_e}
        (-1)^{2F_g}
        \frac{(2F_e+1)^3(2F_g+1)^3}{(2I+1)}
        \sixj(1,1,2)(F_g,F_g,F_e)
        \sixj(J_e,F_e,I)(F_g,J_g,1)^4
        x_{F_g}
        \sbr{L(\Go_{F_eF_g}')}^2\\
        &\times
        \Bigg(
        \frac{(-1)^{2I+2J_g}}{(2F_e+1)(2F_g+1)}
        \sixj(1,1,2)(F_g,F_g,F_e)
        +R\,(-1)^{F_g-F_e}
        (2J_e+1)
        \sixj(1,1,2)(F_e,F_e,F_g)
        \sixj(F_g,F_g,2)(F_e,F_e,1)
        \sixj(J_e,F_e,I)(F_g,J_g,1)^2
        \Bigg).
\end{split}
\end{equation}

\subsection{Doppler-broadened transit effect}
\label{app:dbeffect}

The procedure used to obtain the optical rotation signal in the
Doppler-broadened case is described in
Sec.~\ref{subsec:DBeffect}. When the ground- and excited-state
hyperfine splittings are all much greater than the natural
width ($\Go_{F_eF_e'},\Go_{F_gF_g'},\GG_D\gg\GG$) we have,
applying the integration procedure to
Eq.~\eqref{eq:DFrotationresolved},
\begin{equation}\label{eq:DBrotation}
\begin{split}
    \ell_0\frac{d\Gv}{d\ell}
    &=\frac{3}{8}\Gk_s
        \sum_{F_gF_e}
        (-1)^{2F_g}
        \frac{(2F_e+1)^3(2F_g+1)^3}{(2I+1)}
        \sixj(1,1,2)(F_g,F_g,F_e)
        \sixj(J_e,F_e,I)(F_g,J_g,1)^4
        e^{-(\GD_{F_eF_g}/\GG_D)^2}
        x_{F_g}\\
        &\times
        \Bigg(
        \frac{(-1)^{2I+2J_g}}{(2F_e+1)(2F_g+1)}
        \sixj(1,1,2)(F_g,F_g,F_e)
        +R\,(-1)^{F_g-F_e}
        (2J_e+1)
        \sixj(1,1,2)(F_e,F_e,F_g)
        \sixj(F_g,F_g,2)(F_e,F_e,1)
        \sixj(J_e,F_e,I)(F_g,J_g,1)^2
        \Bigg),
\end{split}
\end{equation}
where the unsaturated absorption length for the
Doppler-broadened case is given by Eq.~\eqref{eq:dbabslength}.

In another limit in which the ground-state hyperfine splittings
are much greater than the natural width and the excited-state
splittings are much less than the Doppler width
($\Go_{F_eF_e'},\GG\ll\GG_D$, $\GG\ll\Go_{F_gF_g'}$), we have
\begin{equation}\label{eq:DBrotsmallAe}
\begin{split}
    \ell_0\frac{d\Gv}{d\ell}
    &=\frac{3}{8}\Gk_s
        \sum_{F_gF_eF_e''}
        (-1)^{2F_g+F_e''-F_e}
        \frac{(2F_e+1)(2F_e''+1)(2F_g+1)^2}{(2I+1)}
        \sixj(1,1,2)(F_g,F_g,F_e'')
        \sixj(J_e,F_e,I)(F_g,J_g,1)^2
        \sixj(J_e,F_e'',I)(F_g,J_g,1)^2
        x_{F_g}\\
    &\qquad\times
        \Bigg(
        (-1)^{2I+2J_g}
        \sixj(1,1,2)(F_g,F_g,F_e)
        +R\sum_{F_e'}
        (-1)^{F_g-F_e'}
        (2J_e+1)
        (2F_g+1)(2F_e'+1)
        \sixj(1,1,2)(F_e,F_e',F_g)
        \sixj(F_g,F_g,2)(F_e,F_e',1)\\
        &\qquad\qquad\qquad\qquad\qquad\qquad\times
        \sixj(J_e,F_e',I)(F_g,J_g,1)^2
        \frac{\sbr{2\GG^4
                +(2\GG^2+\Go^2_{F_eF_e'})
                \Go_{F_eF_e''}\Go_{F_e'F_e''}
                }}
             {2(\GG^2+\Go^2_{F_eF_e'})
              (\GG^2+\Go^2_{F_e'F_e''})}
        \Bigg)
        \frac{e^{-(\GD_{F_g}/\GG_D)^2}\GG^2}{\GG^2+\Go^2_{F_eF_e''}}.
\end{split}
\end{equation}

\subsection{Wall effect}
\label{app:walleffect}

The procedure for obtaining the signal in the wall effect case
is described in Sec.~\ref{subsec:walleffect}. For excited-state
hyperfine splittings much greater than the natural width
($\GG\ll\Go_{F_eF_e'},\GG_D$), we have for the ground-state
polarization
\begin{equation}\label{eq:Wallpol}
\begin{split}
    \Gr_a^{(20)}(F_gF_g)
    &=-\sqrt{\frac{\Gp}{6}}
        \sum_{F_e}
        (-1)^{F_g-F_e}\Gk_s
        \frac{(2F_e+1)(2F_g+1)}{(2I+1)(2J_g+1)}
        \sixj(J_e,F_e,I)(F_g,J_g,1)^2
        \Bigg(
        (-1)^{2I+2J_g}
        \sixj(1,1,2)(F_g,F_g,F_e)
        e^{-(\GD_{F_eF_g}/\GG_D)^2}\\
        &\qquad\qquad
        +R\sum_{F_g'}
        (-1)^{F_g'-F_e}
        (2J_e+1)
        (2F_g'+1)(2F_e+1)
        \sixj(1,1,2)(F_e,F_e,F_g')
        \sixj(F_g,F_g,2)(F_e,F_e,1)
        \sixj(J_e,F_e,I)(F_g',J_g,1)^2
        \Bigg),
\end{split}
\end{equation}
where the saturation parameter for the wall effect is defined
by Eq.~\eqref{eq:wallsatparam}.
The optical rotation signal is then given by
\begin{equation}\label{eq:WallRotation}
\begin{split}
    \ell_0\frac{d\Gv}{d\ell}
    &=-\frac{3\sqrt{3}}{4\sqrt{2}}
        \sum_{F_gF_e'}
        (-1)^{F_g+F_e'}
        (2F_g+1)(2F_e'+1)(2J_g+1)
        \sixj(1,1,2)(F_g,F_g,F_e')
        \sixj(J_g,F_g,I)(F_e',J_e,1)^2
        e^{-(\GD_{F_e'F_g}/\GG_D)^2}
        x_{F_g}\Gr_a^{(20)}(F_gF_g)\\
    &=\frac{3\sqrt{\Gp}}{8}\Gk_s
        \sum_{F_gF_eF_e'}
        (-1)^{2F_g+F_e'-F_e}
        \frac{(2F_e+1)(2F_e'+1)(2F_g+1)^2}{(2I+1)}
        \sixj(1,1,2)(F_g,F_g,F_e')
        \sixj(J_e,F_e,I)(F_g,J_g,1)^2
        \sixj(J_e,F_e',I)(F_g,J_g,1)^2\\
        &\qquad\times
        x_{F_g}
        e^{-(\GD_{F_e'F_g}/\GG_D)^2}\Bigg(
        (-1)^{2I+2J_g}
        \sixj(1,1,2)(F_g,F_g,F_e)
        e^{-(\GD_{F_eF_g}/\GG_D)^2}\\
        &\qquad\quad+R\sum_{F_g'}
        (-1)^{F_g'-F_e}
        (2J_e+1)(2F_e+1)(2F_g'+1)
        \sixj(1,1,2)(F_e,F_e,F_g')
        \sixj(F_g,F_g,2)(F_e,F_e,1)
        \sixj(J_e,F_e,I)(F_g',J_g,1)^2
        e^{-(\GD_{F_eF_g'}/\GG_D)^2}
        \Bigg).
\end{split}
\end{equation}
\end{widetext}

\bibliography{Simons_NMORBib,Atom_Light_Int}

\end{document}